\documentclass[prd,showpacs,showkeys,preprintnumbers,floatfix,groupedaddress,nofootinbib,superscriptaddress,11pt]{revtex4-1}
\usepackage{hyperref} 
\usepackage{graphicx}
\usepackage{amssymb}
\usepackage{amsmath}
\usepackage{bbold}
\usepackage{epstopdf}
\usepackage{float}
\usepackage{caption}
\usepackage{slashed}
\usepackage{color}
\usepackage{verbatim}
\usepackage{multirow}
\usepackage[singlelinecheck=false]{subcaption}

\DeclareGraphicsExtensions{.pdf,.png,.jpg}
\DeclareMathOperator{\tr}{tr}
\def\chpt{\raise0.4ex\hbox{$\chi$}{\rm PT}}
\def\lchpt{L\raise0.4ex\hbox{$\chi$}PT}

\def\QCD{{\rm QCD}}
\newcommand{\MSb}{\overline{\textrm{MS}}}

\begin{document}

\title{Topological susceptibility from twisted mass fermions\\ using spectral projectors and the gradient flow}

\author{Constantia Alexandrou}
\affiliation{
Department of Physics, University of Cyprus,
P.O. Box 20537, 1678 Nicosia, Cyprus \\
}
\affiliation{
Computation-based Science and Technology Research Center,
Cyprus Institute, 20 Kavafi Str., Nicosia 2121, Cyprus\\
}
\author{Andreas Athenodorou}
\affiliation{
Computation-based Science and Technology Research Center,
Cyprus Institute, 20 Kavafi Str., Nicosia 2121, Cyprus\\
}
\author{Krzysztof Cichy}
\email[email: ]{kcichy@amu.edu.pl}
\affiliation{
Goethe-Universit\"at Frankfurt am Main, Institut f\"ur Theoretische Physik,
Max-von-Laue-Strasse 1, 60438 Frankfurt am Main, Germany \\
}
\affiliation{
Faculty of Physics, Adam Mickiewicz University,
Umultowska 85, 61-614 Pozna\'n, Poland \\
}
\author{Martha Constantinou}
%
%
\affiliation{
 Physics Department, Temple University, 
 Philadelphia, PA 19122-1801, USA \\
}
\author{Derek P. Horkel}
\email[e-mail: ]{derek.horkel@temple.edu}
\affiliation{
 Physics Department, Temple University, 
 Philadelphia, PA 19122-1801, USA \\
}

\author{Karl Jansen}
\affiliation{
NIC, DESY, Platanenallee 6, D-15738 Zeuthen, Germany\\
}

\author{Giannis Koutsou}
\affiliation{
Computation-based Science and Technology Research Center,
Cyprus Institute, 20 Kavafi Str., Nicosia 2121, Cyprus\\
}

\author{Conor Larkin}
\affiliation{
Computation-based Science and Technology Research Center,
Cyprus Institute, 20 Kavafi Str., Nicosia 2121, Cyprus\\
}

\date{\today}

\begin{abstract}
\begin{center}
\includegraphics
[width=0.2\textwidth,angle=0]
{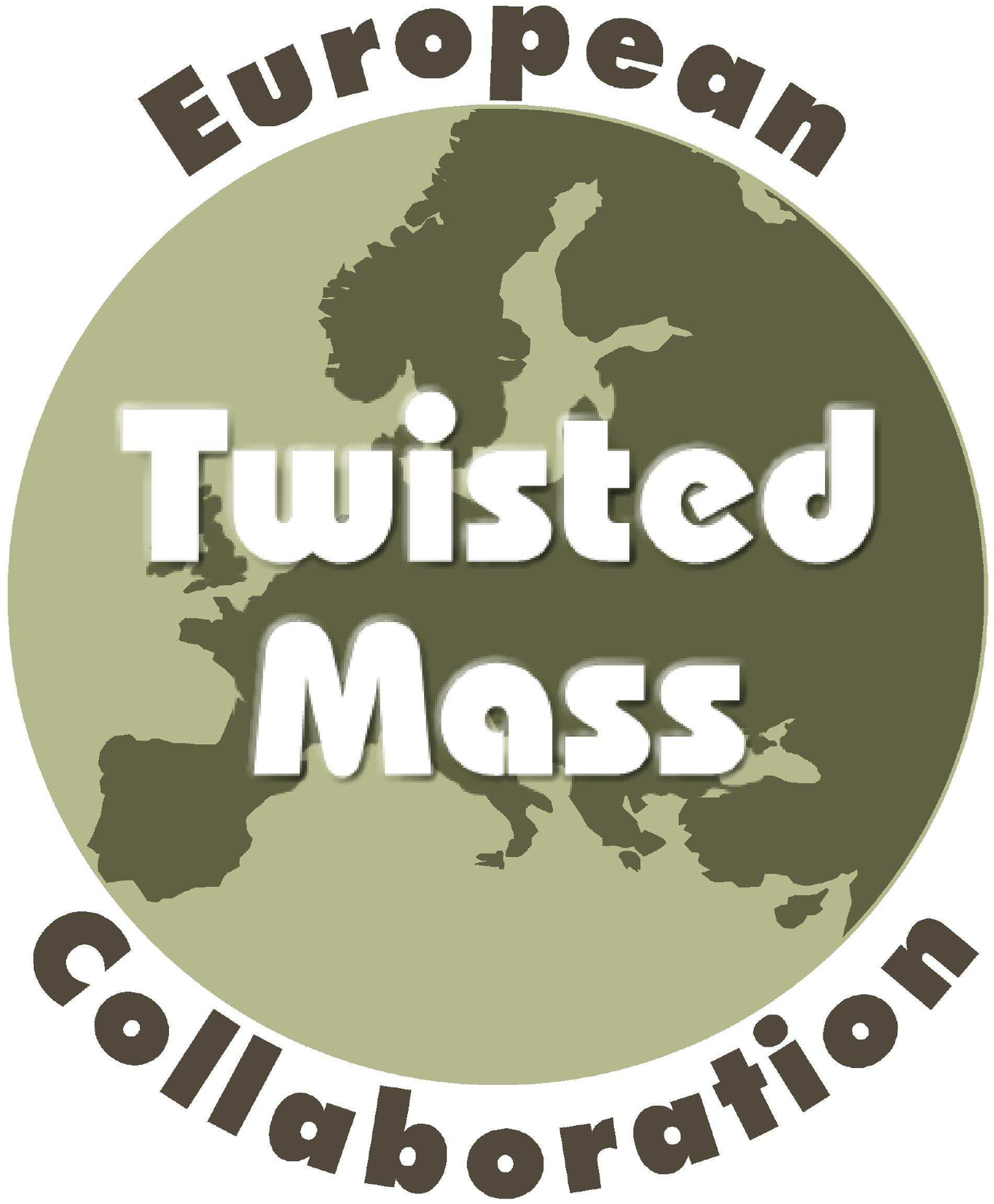}
\end{center}
We compare lattice QCD determinations of topological susceptibility using a gluonic definition from the gradient flow and a fermionic definition from the spectral projector method.
We use ensembles with dynamical light, strange and charm flavors of maximally twisted mass fermions.
For both definitions of the susceptibility we employ ensembles  at three values of the lattice spacing and several quark masses at each spacing. 
The data are fitted to chiral perturbation theory predictions with a discretization term to determine the continuum chiral condensate in the massless limit and estimate the overall discretization errors.
We find that both approaches lead to compatible results in the continuum limit, but the gluonic ones are much more affected by cut-off effects.
This finally yields a much smaller total error in the spectral projector results.
We show that there exists, in principle, a value of the spectral cutoff which would completely eliminate discretization effects in the topological susceptibility.
\end{abstract}

\maketitle

\section{Introduction}
\label{sec:intro}

Topology is essential in understanding the physics of QCD.
Due to the fact that the lattice inherently has different topology from the continuum, an accurate and precise measurement of topological charge is a difficult task in lattice QCD, with a long history of failed attempts and practical and theoretical issues to overcome \cite{Luscher:2004fu,Luscher:2010ik,Cichy:2014qta,Muller-Preussker:2015daa,Alexandrou:2015yba,Alexandrou:2017hqw}.
Yet, a good definition of topological charge is essential to modern lattice calculations, e.g.\ ones involving CP-violation. In particular, the neutron electric dipole moment (nEDM) depends on a topological charge dependent parameter called $\alpha^1$, defined in Eq.~(31) of \cite{Alexandrou:2015spa}. A significant source of error in the nEDM calculation and the calculation of the mass of the $\eta$ meson, is due to uncertainty in the topological charge \cite{Alexandrou:2015ttm}. 

There are several definitions of topological charge that belong to two broad classes, fermionic and gluonic definitions.
In the continuum, they are all equivalent, but on the lattice they can have very different discretization effects.
The gluonic definition, utilizing various smearing techniques to damp UV fluctuations, such as the gradient flow (GF) \cite{Luscher:2010iy,Luscher:2011bx,Luscher:2013vga} that we use in this paper, is computationally relatively cheap to calculate and is currently widely used, for example, in the aforementioned nEDM calculation \cite{Alexandrou:2015ttm}. 
In this paper we focus on a fermionic definition using the method of spectral projectors \cite{Giusti:2008vb,Luscher:2010ik}, but we also compare the results to ones from a GF-smeared gluonic definition.

Given the topological charge, $Q$, on every gauge field configuration, the topological susceptibility, $\chi$, is given by the ensemble average $\langle Q^2\rangle$, divided by the lattice volume $V$. 
Hence, it is essentially a quantity which provides a measure of topological charge fluctuations.
Using leading order chiral perturbation theory ($\chpt$) and parametrizing the leading discretization term, the topological susceptibility can be written as a simple equation depending on the quark mass and lattice spacing. 

In this paper, the coefficient of the mass term, which is a low energy constant (LEC) known as the chiral condensate, $\Sigma$, is also extracted, from both the spectral projector definition and the GF-smeared gluonic definition.
Our aim is, in particular, a quantitative investigation of cut-off effects present in both formulations, motivated by our observations that the gluonic topological charge is subject to large discretization effects, at least in the setup of our lattice study (see below).

This paper is organized as follows: Section \ref{sec:setup} presents the lattice setup. In Section \ref{sec:theory}, we briefly review the spectral projector method and the gluonic definition of the topological charge (with GF as the smoother), as well as the $\chpt$ formula for topological susceptibility. Section \ref{sec:extrap} contains results from the two methods, together with fits used to determine the continuum value of the chiral condensate. Lastly, we conclude in Section \ref{sec:conclusions} reviewing our results and discussing future directions.

\section{Lattice setup}
\label{sec:setup}

The following calculations were performed using twisted mass configurations with $N_f=2+1+1$ dynamical flavors provided by the European Twisted Mass Collaboration (ETMC)  \cite{Baron:2010bv,Baron:2010th,Baron:2011sf}.
The Iwasaki action \cite{Iwasaki:1985we} is used in the gauge sector. 
The action in the fermionic sector for the light flavors (in the twisted basis) reads: \cite{Frezzotti:2000nk,Frezzotti:2003xj,Frezzotti:2003ni,Frezzotti:2004wz}
\begin{equation}
 S_l = a^4 \sum_x \bar{\chi}_l(x) \left( D_W + m_0 + i \mu_l \gamma_5 \tau_3
\right)
\chi_l(x),
\end{equation}
where $\tau^3$ is the third Pauli matrix, acting in flavor space, $\chi_l=(\chi_u, \chi_d)$ is a vector in flavor space and $D_W$ is the standard Wilson-Dirac operator.
The bare mass parameters, $m_0$ and $\mu_l$, are the untwisted and twisted light quark masses, respectively. 
The renormalized light quark mass, which we denote by $\mu_{l,R}$, is related to the bare twisted mass $\mu_{l,R}=Z_P^{-1}\mu_l$, where $Z_P$ is the renormalization function of the pseudoscalar density. 

The action for the heavy flavors is: \cite{Frezzotti:2003xj,Frezzotti:2004wz}
\begin{equation}
 S_h = a^4 \sum_x \bar{\chi}_h(x) \left( D_W + m_0 + i \mu_\sigma \gamma_5
\tau_1 + \mu_\delta \tau_3
\right)
\chi_h(x),
\end{equation}
where $\mu_\sigma$ is the bare twisted mass with the twist along the $\tau_1$ direction in charm-strange flavor space
and $\mu_\delta$ gives the splitting along the $\tau_3$ direction, i.e.\ generates unequal strange and charm quark masses.
$\chi_h=(\chi_c,\,\chi_s)$ is a vector in charm-strange flavor space, written in the twisted basis.

Twisted mass fermions have the advantage of automatic $\mathcal{O}(a)$ improvement when tuned to maximal twist. Maximal twist is obtained by tuning the hopping parameter, $\kappa=(8 + 2am_0)^{-1}$, to its critical value $\kappa_c$, where the PCAC mass vanishes \cite{Frezzotti:2003ni,Farchioni:2004fs,Frezzotti:2005gi,Jansen:2005kk,Chiarappa:2006ae}.

Three different lattice spacings were used, ranging from $\sim0.062-0.089$ fm \cite{Carrasco:2014cwa}. 
For each spacing, we performed calculations for three renormalized quark masses, spanning the interval $12-28$ MeV. 
Topological charge was measured using the gluonic definition on $\mathcal{O}(1000)$ configurations and using the method of spectral projectors on $\mathcal{O}(100-200)$ configurations. 
Both sets of configurations cover the same range in Monte Carlo time, but the spectral projector method was used on fewer configurations due to it having greater computational demand. The lattice parameters are collected in Tables~\ref{tab:latticeparams1} 
and \ref{tab:latticeparams2}, while in Table~\ref{tab:MC} we give information on the measurements.

\begin{table}
   \centering
  \begin{tabular}{cccccccccccc}
\hline  
\hline  
    Ensemble & $\beta$ & Lattice size & $a\mu_l$ & $\mu_{l,R}$ [MeV]&
    $\kappa_c$ & $N_{\rm GF}$ & $N_{\rm spec}$ \\
\hline
A30.32 &1.90 & $32^3\times 64$  & 0.0030 & 12.6 & 0.163272 & 1100 & 224\\
A40.24 &1.90 & $24^3\times 48$  & 0.0040 & 16.8 & 0.163270  & 1125 & 180\\
A60.24 &1.90 & $24^3\times 48$  & 0.0060 & 25.1 & 0.163265  & 1160 & 212\\
\bf A80.24 & \bf 1.90 & $\bf 24^3\times 48$  & \bf  0.0080 & \bf  33.5 & \bf  0.163260 & \bf  1100 & -\\
\bf A100.24 & \bf 1.90 & $\bf 24^3\times 48$  & \bf  0.0100 & \bf  41.9 & \bf  0.163255 & \bf  1018 & -\\[1ex]
B25.32 & 1.95 & $32^3\times64$  & 0.0025 & 11.8& 0.161240 & 971 & 248  \\
B35.32 &1.95 & $32^3\times64$  & 0.0035 & 16.5& 0.161240 & 1000 & 201 \\
B55.32 & 1.95 & $32^3\times64$  & 0.0055 &26.0& 0.161236 & 4689 & 470  \\
\bf B75.32 & \bf  1.95 & $\bf 32^3\times64$  & \bf  0.0075 & \bf 35.5& \bf  0.161232 & \bf  559 & -\\
\bf B85.24 & \bf  1.95 & $\bf 24^3\times48$  & \bf  0.0085 & \bf 40.2& \bf  0.161231 & \bf  1128 & -\\[1ex]
\bf D15.48 & \bf   2.10 & $\bf 48^3\times96$  & \bf  0.0015 & \bf  9.3& \bf 0.156361& \bf  646 & -\\
D20.48 &  2.10 & $48^3\times96$  & 0.0020 & 12.3 &0.156357& 1429 & 94 \\
D30.48 & 2.10 & $48^3\times96$  & 0.0030 & 18.5 &0.156355  & 1947 & 102\\
D45.32 & 2.10 & $32^3\times64$  & 0.0045 & 27.8 & 0.156315  & 949 & 95\\
\hline  
\hline  
\end{tabular}
  \caption{{\small{Parameters of gauge field ensembles, taken from
\cite{Baron:2010bv,Baron:2010th,Baron:2011sf,Ottnad:2012fv,ETM:2011aa,Carrasco:2014cwa,Alexandrou:2015spa}. 
Shown parameters are: the inverse bare coupling $\beta$,
lattice size $(L/a)^3\times(T/a)$, bare twisted light quark mass $\mu_l$,
renormalized quark mass $\mu_{l,R}$, critical value of the hopping parameter, and the number of configurations topological charge was measured on using the gluonic definition ($N_{\rm GF}$) and the spectral projector method ($N_{\rm spec}$). Bolded ensembles were analyzed using only the gluonic definition. In order to have a meaningful comparison, these ensembles are not used in any fits, but they are shown in Fig. \ref{fig:paramGF}.}}}
  \label{tab:latticeparams1}
\end{table}

\begin{table}
   \centering
  \begin{tabular}{cccccc}
  \hline  
  \hline  
  $\beta$ & $a$ [fm] & $Z_P$& $Z_P/Z_S$ & $r_0/a$ \\
\hline
1.90 &  0.0885(36) & 0.529(9) & 0.699(13)& 5.231(38)\\
1.95 &  0.0815(30) & 0.504(5) & 0.697(7)& 5.710(41)\\
2.10 &  0.0619(18) & 0.514(3) & 0.740(5)& 7.538(58)\\
\hline  
\hline  
\end{tabular}
  \caption{{\small{Parameters of gauge field ensembles, common for the same value of $\beta$, taken from
\cite{Baron:2010bv,Baron:2010th,Baron:2011sf,Ottnad:2012fv,ETM:2011aa,Carrasco:2014cwa,Alexandrou:2015spa}. 
Shown parameters are: $\beta$,
the lattice spacing $a$, the Sommer parameter in lattice units $r_0/a$, the scheme- and
scale-independent renormalization functions ratio $Z_P/Z_S$ of the pseudoscalar and scalar densities, and the renormalization function $Z_P$ of the pseudoscalar density in
the $\MSb$ scheme at the scale of 2 GeV.}}}
  \label{tab:latticeparams2}
\end{table}

\begin{table}
   \centering
  \begin{tabular}{ccccccc}
  \hline  
  \hline  
    \multirow{2}*{Ensemble} &\multirow{2}*{\rm $\rm step_{spec}$}& \multirow{2}*{$\tau_{\rm int,spec}$} & \multirow{2}*{$\rm step_{GF}$} &\multirow{2}*{$\tau_{\rm int,GF}$}& \multicolumn{2}{c}{Relative errors}\\
     &&&& & spec & GF\\
\hline
A30.32 &24&0.58&4&1.09 &  10\%&  7\% \\
A40.24 &20&0.50&4&0.92 &  11\% &  6\%\\
A60.24 &20&0.53&4&0.73 &  10\%&  5\% \\
\bf A80.24 &-&-&\bf 4&\bf 0.79 &-&\bf  5\%\\
\bf A100.24 &-&-&\bf 4&\bf 0.75&-&\bf  5\%\\ [1ex]
B25.32 & 24&0.49&4&0.72 &  8\%& 6\%\\
B35.32 & 20&0.43&4&1.01 &  11\%  & 6\%\\
B55.32 &20&0.69&2&1.65 &  8\% & 4\%\\
\bf B75.32 &-&-& \bf 4&\bf 0.53 &-&\bf  6\%\\
\bf B85.24 &-& -& \bf 4&\bf 0.80 &-&\bf  6\%\\ [1ex]
\bf D15.48 &-& - & \bf 2&\bf 1.86 &-&\bf  12\%\\
D20.48 &40&0.67&2&2.20 & 18\% &  8\% \\
D30.48 &40&0.60&2&5.11 & 15\% &  11\%\\
D45.32 &40&0.41&4&2.19 & 13\% &  11\%\\

\hline
\hline
\end{tabular}
  \caption{{\small{Ensemble label, the step between measurements (in units of molecular dynamics trajectories), integrated autocorrelation time and the relative error of the extracted topological susceptibility. The spectral projector quantities regard the case of $M=120$ MeV and are denoted by ``spec'', while the gluonic ones are denoted by ``GF''. Bolded ensembles were only analyzed using the gluonic definition. }}}
  \label{tab:MC}
\end{table}

\section{Theory}
\label{sec:theory}

The method of spectral projectors was introduced and described in detail in Refs.~\cite{Giusti:2008vb,Luscher:2010ik}.
In these papers, the projector was constructed stochastically, thus limiting the computational cost with respect to an explicit computation of eigenmodes from $\mathcal{O}(V^2)$ to $\mathcal{O}(V)$, at the expense of introducing stochastic noise.
For the current investigation, we opted for an explicit computation of eigenmodes, in order not to have this stochastic contamination.
In practice, we used the ARPACK (ARnoldi PACKage) eigensolver library \cite{doi:10.1137/1.9780898719628} to calculate the lowest 400 eigenmodes of the normal massless Wilson-Dirac operator, $D_{W}^\dagger D_{W}$. We performed the calculations on P100 Nvidia GPU architecture, using the ETMC's code library \cite{ETMCGit}, which uses the QUDA linear solver library as its base \cite{Clark:2009wm}.

Here, we summarize the relevant details for calculating the topological susceptibility from the Hermitian Wilson-Dirac operator ($D_{W}^\dagger D_{W}$) spectrum. The bare topological charge can be defined as
\begin{equation}
Q_0=\sum_i^{\lambda_i < M_0^2} R_i,~~R_i= u_i^\dagger \gamma_5 u_i
\end{equation}
and the bare topological susceptibility as
\begin{equation}
\chi_{0}= \frac{\langle Q_0^2\rangle}{V}.
\label{eqn:specproj}
\end{equation}
The renormalized quantities are
\begin{equation}
\chi= \left(\frac{Z_S}{Z_P}\right)^2 \chi_0,~~M=Z_P^{-1} M_0.
\end{equation}
In the above formulae, $u_i$ are eigenvectors of the normal operator $D_{W}^\dagger D_{W}$ 
and $M^2$ is the renormalized spectral threshold, i.e.\ all the renormalized eigenvalues taken into account have magnitude smaller than $M^2$. The 400 lowest eigenmodes allow $M$ up to $160$ MeV for all ensembles described in Table~\ref{tab:latticeparams1}.

For chiral fermions, such as overlap fermions, the topological charge receives contributions only from exact zero modes of $D_{W}^\dagger D_{W}$. The zero modes correspond to $\lambda=0$, such that $R_{\lambda=0} = \pm 1$ and $R_{\lambda>0} = 0$.
This would indicate that a spectral threshold slightly larger than zero would be enough to obtain a topological charge exactly equivalent to the one from the index of the overlap Dirac operator \cite{Neuberger:1997fp,Neuberger:1998wv,Niedermayer:1998bi}.
For non-chiral fermions, such as Wilson twisted mass fermions which we use in this paper, the would-be zero and non-zero modes are shifted by $\mathcal{O}(a^2)$ effects. 
Moreover, different values of the spectral threshold may correspond to different cut-off effects.
An optimal value of $M$, i.e.\ one minimizing discretization effects, can be chosen by analyzing several values of the lattice spacing.
This allows one to find such a value that the slope of the topological susceptibility vs.\ the lattice spacing is minimal, as we show explicitly below.
The value that we find is appropriate for the here considered setup and it may be different when using e.g.\ a different fermion discretization.
If one works with a single ensemble, it is, of course, not possible to find such an optimal value.
However, one can still check the dependence of the topological susceptibility on $M$ and if a region of small variation of the susceptibility is found, it likely corresponds to small cut-off effects, as we also observe in our data.
Note that the topological susceptibility from the spectral projector method is automatically $\mathcal{O}(a)$-improved when using twisted mass fermions at maximal twist \cite{Cichy:2013rra}.

The other definition of the topological charge that we employ in this paper is the gluonic (field theoretic) one.
The calculation of the topological charge with this definition proceeded along the lines of Ref.~\cite{Alexandrou:2015yba}, which we refer to for more details. 
The field theoretic definition reads:
\begin{equation}
q(x)=\frac{1}{32\pi^2} \epsilon_{\mu\nu\rho\sigma} \tr\left[G_{\mu\nu}(x) G_{\rho\sigma}(x)\right],~~~~ Q=a^4 \sum_x q(x) ,  ~~~~ \chi = \frac{\langle{Q^2\rangle}}{V}.
\label{eqn:gfdef}
\end{equation}
The definition of the field strength tensor used is tree-level Symanzik improved, i.e.\ $\mathcal{O}(a^2)$ improved
by including both clover terms and rectangular Wilson loops, as described in Ref.~\cite{Alexandrou:2015yba}. 
UV fluctuations  are filtered using the GF procedure, as described in Ref.~\cite{Luscher:2010iy}.
The smoothing action in the GF is the tree-level Symanzik improved gauge action.

Next, we derive an expression for the topological susceptibility in chiral perturbation theory, following the arguments used in 
Ref.~\cite{Mao:2009sy} but adapted for twisted mass. 
First, we consider the partition function of QCD with non-zero $\Theta_\QCD$:
\begin{equation}
Z(\Theta)=\sum_Q \int \! \mathcal{D}A_\mu \mathcal{D} \psi \mathcal{D}\psi^\dagger e^{i S_{\QCD,\Theta=0} -i \Theta Q[U]}.
\end{equation}
The topological susceptibility can be written as the second derivative of the energy density, evaluated at $\Theta=0$,
\begin{equation}
\chi = \left.\frac{\partial^2 \varepsilon(\Theta)}{\partial \Theta^2} \right\vert_{\Theta=0} = \frac{\langle{Q^2\rangle}}{V},~~~~\varepsilon(\Theta) = -\frac{\log(Z)}{V}~.
\label{eqn:chiQCD}
\end{equation}
The lowest order, two flavor $\chpt$ Lagrangian can be written as:
\begin{equation}
\mathcal{L}_{\chpt}^{(2)} = \frac{f_\pi ^2}{4} \tr\left[ \partial_\mu U(x) \partial^\mu U(x)^\dagger \right] + \Sigma\, {\rm Re}\left(\tr\left[ \mathcal{M} U(x)^\dagger \right]\right),
\end{equation}
where $U(x)$ is an element of $SU$(2) that can be expressed as $U(x)= \mathbb{1}\cos(\phi(x)) + i \hat{n} \cdot \vec{\tau} \sin(\phi(x))$, where $\hat{n}$ is a unit vector in $\mathfrak{su}$(2) flavor space. $\mathcal{M}$ is the quark mass matrix, which for two maximally twisted light flavors is just $\mathcal{M} = i \tau_3 \mu_l$. $f_\pi$ and $\Sigma$ are low energy constants, identified as the pion decay constant and the chiral condensate, respectively.
$\Theta_\QCD$ can be included by using the fact that the physical vacuum angle depends on the argument of the determinant of the quark mass matrix, $\Theta_{phys} = \Theta + \arg \det(\mathcal{M})$. 
Thus, a non-zero $\Theta_\QCD$ can be included by rotating it into the mass matrix, $\mathcal{M}\rightarrow \mathcal{M} \tau_3 e^{i \Theta/2}$.

The partition function can now be written in terms of $\mathcal{L}_{\chpt}^{(2)}$:
\begin{equation}
Z(\Theta)= \int \! \mathcal{D}U e^{i \int \! d^4 x \mathcal{L}^{(2)}(U(x),\Theta)}.
\end{equation}
For volumes that are large relative to the quark mass, $V \mu_l \Sigma \gg1$, the group integral is dominated by $U(x)$ and by extension $\phi(x)$, which is $x$ independent and minimizes $-\mathcal{L}_{\chpt}^{(2)}$. 
For maximal twist and non-zero $\Theta$, this minimum occurs at $\phi=\pm \pi/2$, $\hat{n}=(0,0,1)$. The partition function simplifies to:
\begin{equation}
Z(\Theta)=Z_0 e^{2 V \Sigma \mu_l \cos(\Theta/2)},
\end{equation}
which yields:
\begin{equation}
\chi = \left.\frac{\partial^2 \varepsilon(\Theta)}{\partial \Theta^2} \right\vert_{\Theta=0} = \Sigma \frac{\mu_l}{2},
\label{eqn:contchi}
\end{equation}
which is correct to leading order in the quark mass. In the lattice calculation, there will be corrections of order $a^2$, as maximal twist removes any $\mathcal{O}(a)$ effects.

\section{Extrapolation of the chiral condensate}
\label{sec:extrap}

\subsection{Three-parameter fit}

\begin{figure}[t!]
   \centering
   \begin{subfigure}[t]{0.49\textwidth}
   \centering
   \includegraphics[width=\textwidth]{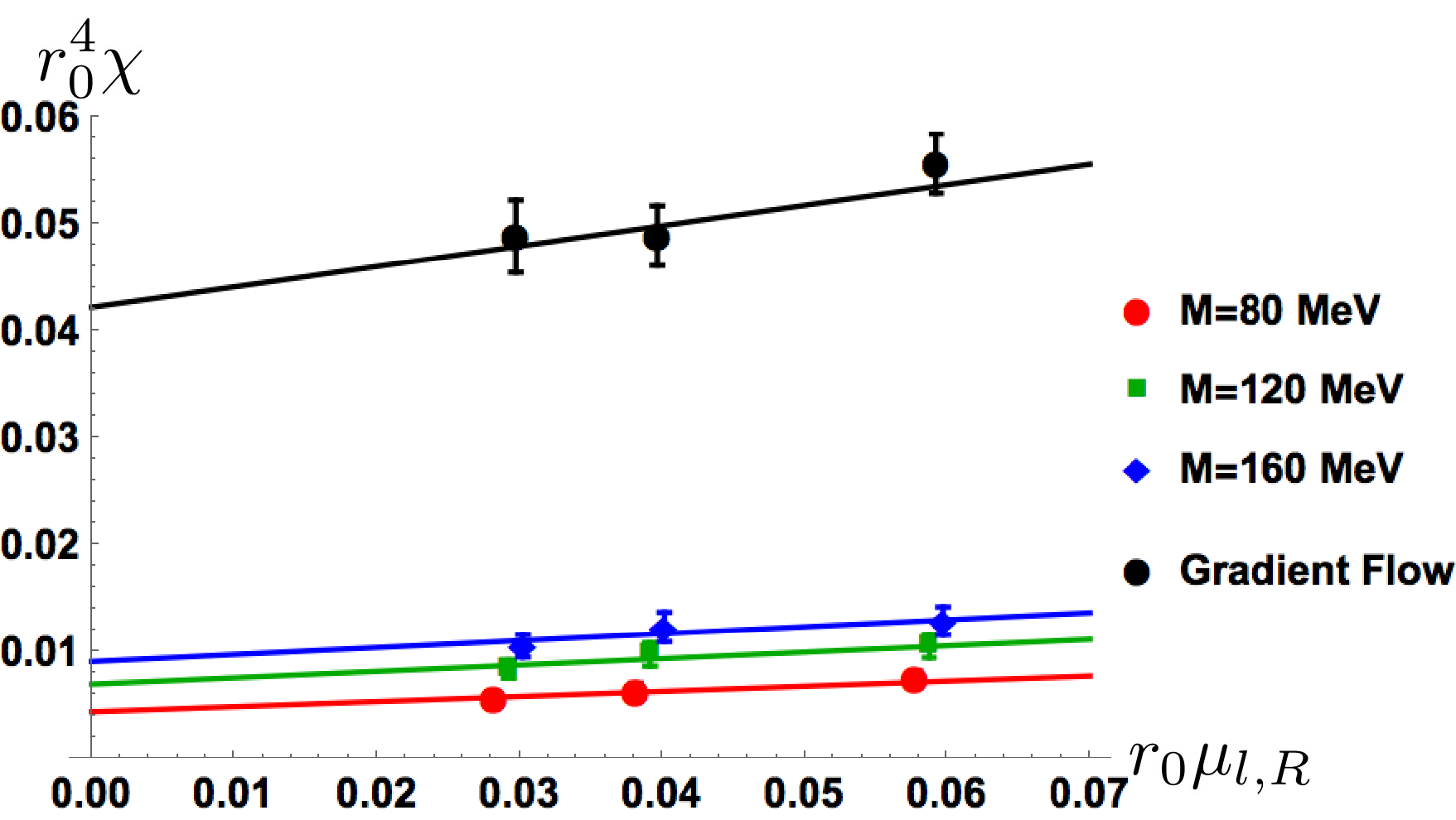}
   \caption{}
   \label{fig:paramA}
\end{subfigure}
   \begin{subfigure}[t]{0.49\textwidth}
   \centering
   \includegraphics[width=\textwidth]{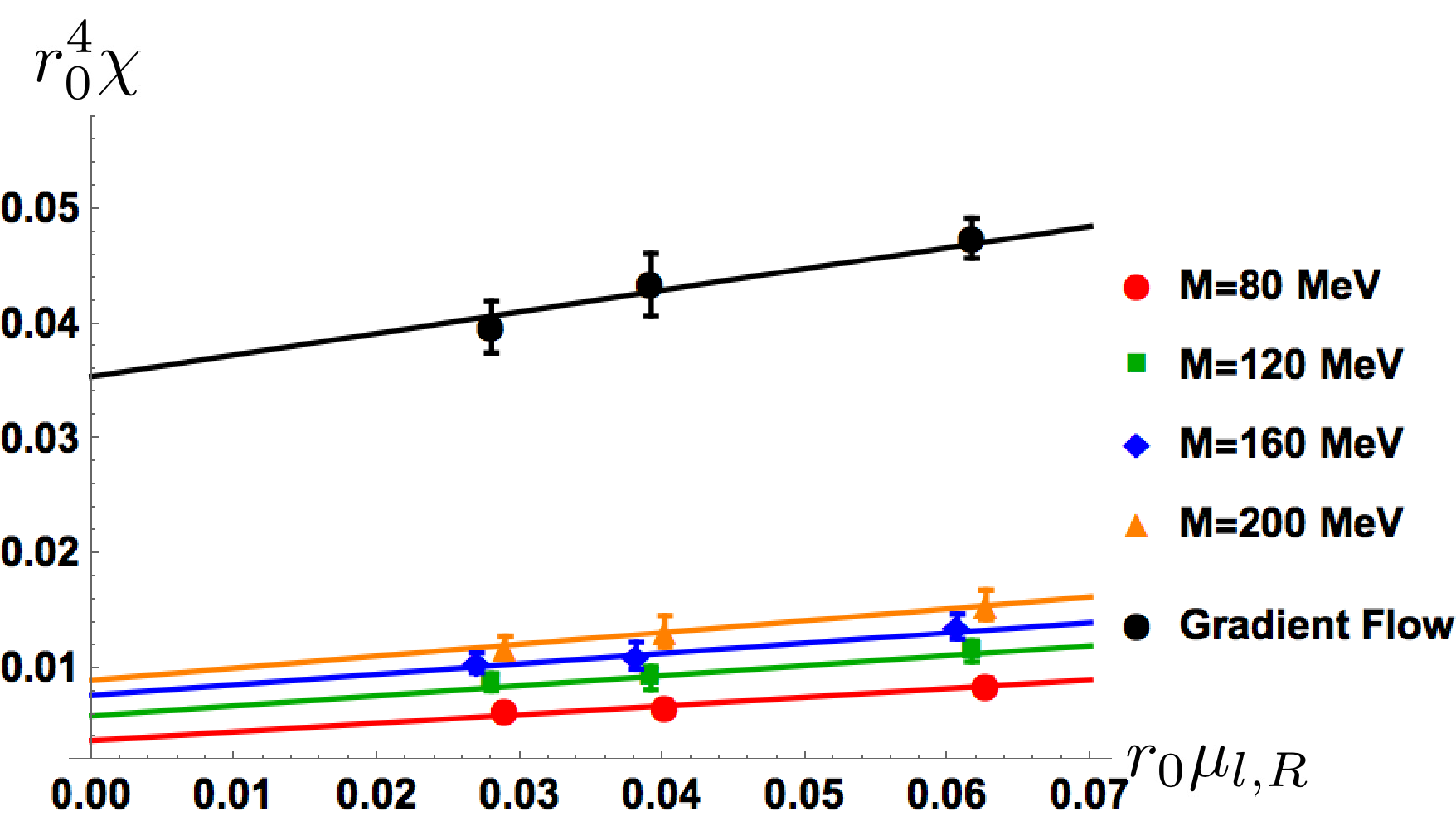}
   \caption{}
   \label{fig:paramB}
   \end{subfigure}
   \begin{subfigure}[t]{0.49\textwidth}
   \centering
   \includegraphics[width=\textwidth]{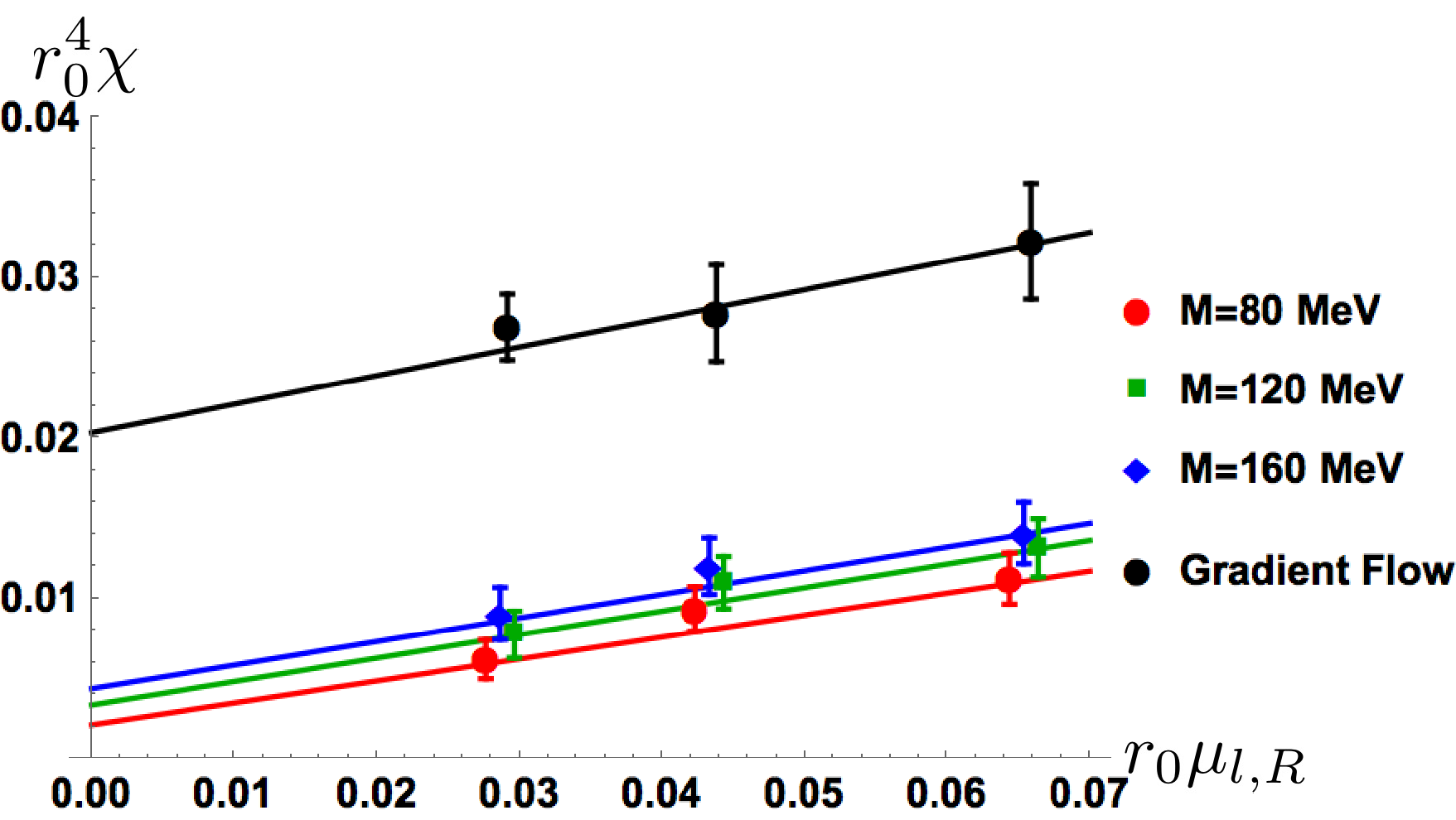}
   \caption{}
   \label{fig:paramD}
   \end{subfigure}
      \caption{\small{Cross sections of global fits of topological susceptibility to the form of Eq.~(\ref{eqn:paramfit}) at $a=0.089,\,0.082,\,0.062$ fm, shown in plots (a), (b) and (c), respectively. 
      Statistical errors are the result of a bootstrap procedure with 1000 samples and appropriate blocking.
      Note that the values are offset on the horizontal axis for visibility. Ensembles used are summarized in Table~\ref{tab:latticeparams1}. 
      Resulting fit parameters are shown in Table~\ref{tab:paramfit}.}}
\end{figure}

\begin{figure}[t!]
   \centering
    \begin{subfigure}[t]{0.49\textwidth}
   \centering
   \includegraphics[width=\textwidth]{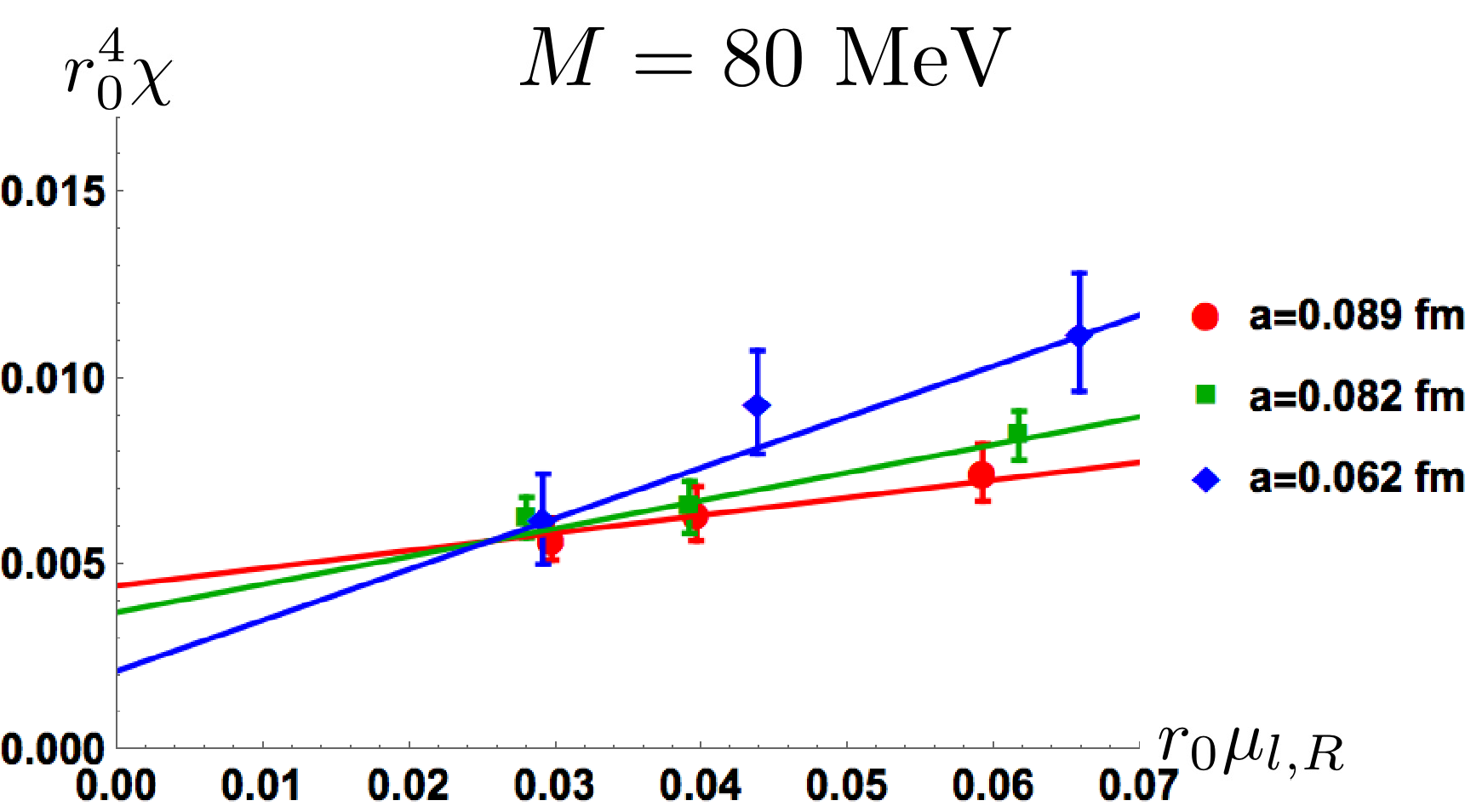}
   \caption{}
   \label{fig:param80}
\end{subfigure}
 \begin{subfigure}[t]{0.49\textwidth}
   \centering
   \includegraphics[width=\textwidth]{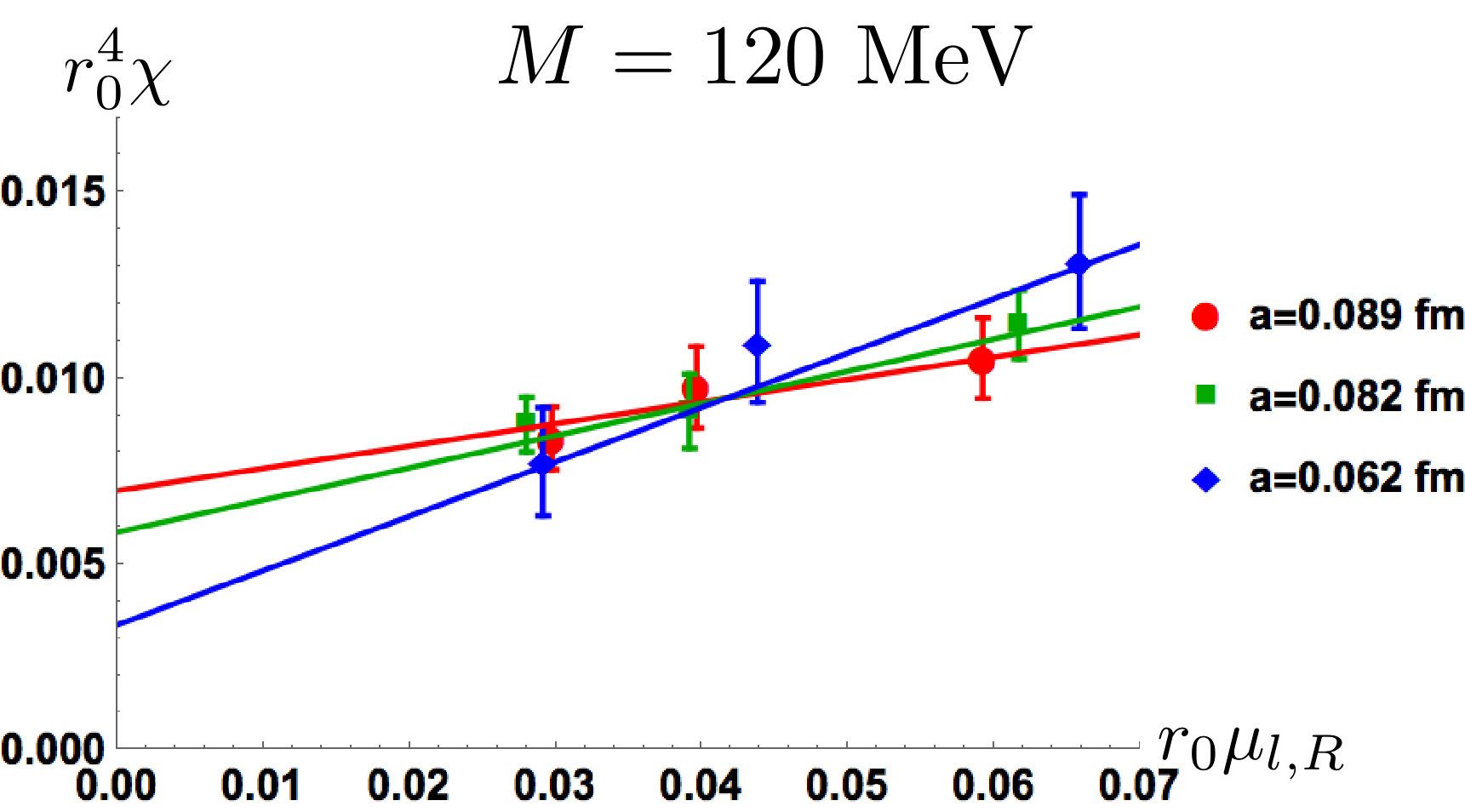}
   \caption{}
   \label{fig:param120}
\end{subfigure}
 \begin{subfigure}[t]{0.49\textwidth}
   \centering
   \includegraphics[width=\textwidth]{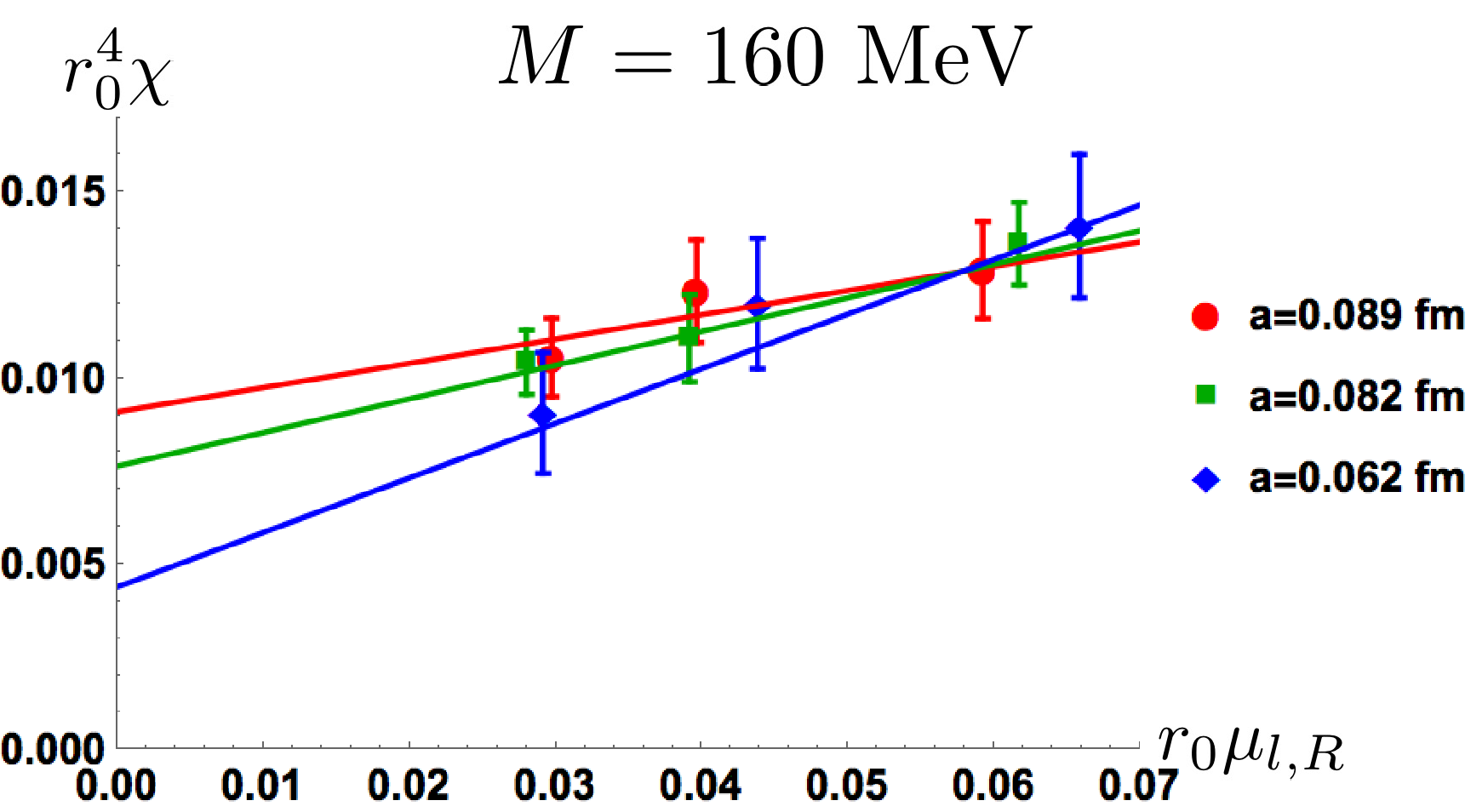}
   \caption{}
   \label{fig:param160}
\end{subfigure}
 \begin{subfigure}[t]{0.49\textwidth}
   \centering
   \includegraphics[width=\textwidth]{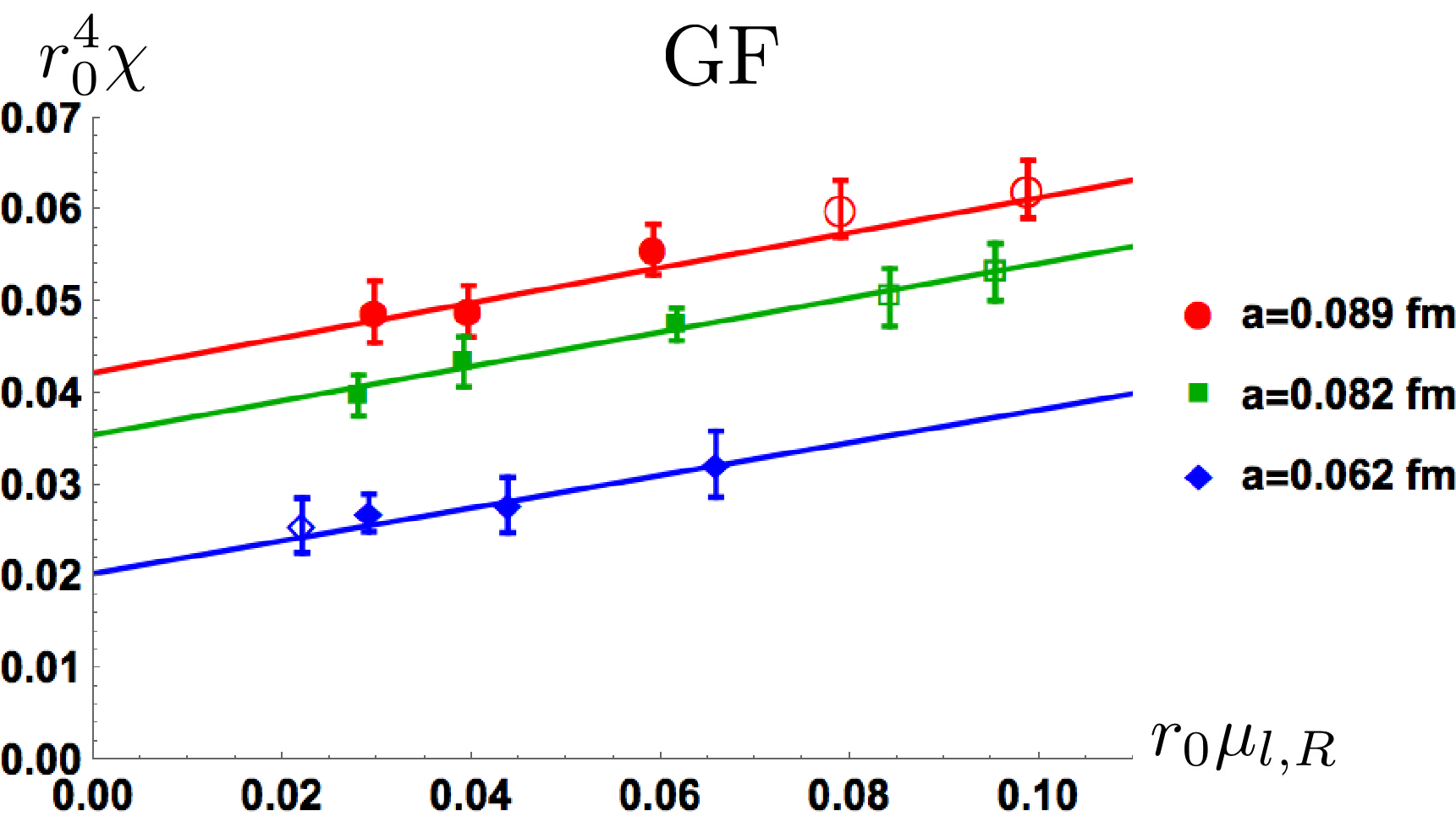}
   \caption{}
   \label{fig:paramGF}
   \end{subfigure}
    \caption{\small{Global fits of topological susceptibility to the form of Eq.~(\ref{eqn:paramfit}) for $M=80, 120, 160$ MeV and the gluonic definition using gradient flow (GF). 
    Statistical errors are the result of a bootstrap procedure with 1000 samples and appropriate blocking.
    Ensembles used are summarized in Table~\ref{tab:latticeparams1}. 
    Resulting fit parameters are shown in Table~\ref{tab:paramfit}. 
    The open points in \ref{fig:paramGF} were not included in the fit to make a direct comparison of spectral projectors and GF on the same ensembles.}}
\end{figure}

\begin{figure}[t!]
   \centering
   \includegraphics[width=.49\textwidth]{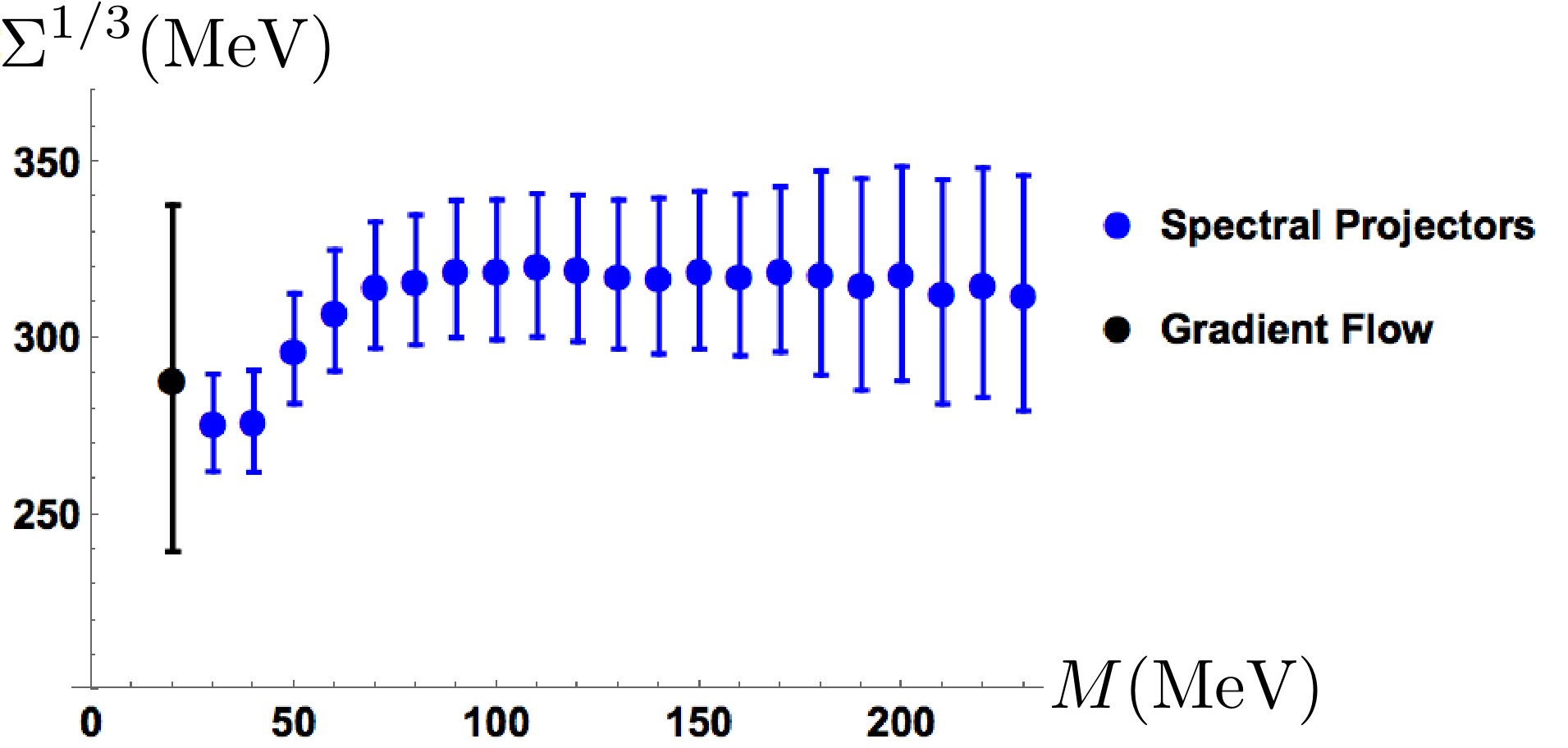}
   \caption{\small{Values of $r_0 \Sigma^{1/3}$ from global fits of topological susceptibility to the form of Eq.~(\ref{eqn:paramfit}). Blue (black) points correspond to spectral projectors (GF).
   The position of the black point along the horizontal axis is arbitrary and unrelated to the scale. 
   Statistical errors are the result of a bootstrap procedure with 1000 samples and appropriate blocking. 
   Ensembles used are summarized in Table~\ref{tab:latticeparams1}. 
   Resulting fit parameters are shown in Table~\ref{tab:paramfit}.}}
   \label{fig:paramSig}
\end{figure}

\begin{figure}[htb!]
   \centering
   \begin{subfigure}[t]{0.49\textwidth}
   \centering
   \includegraphics[width=\textwidth]{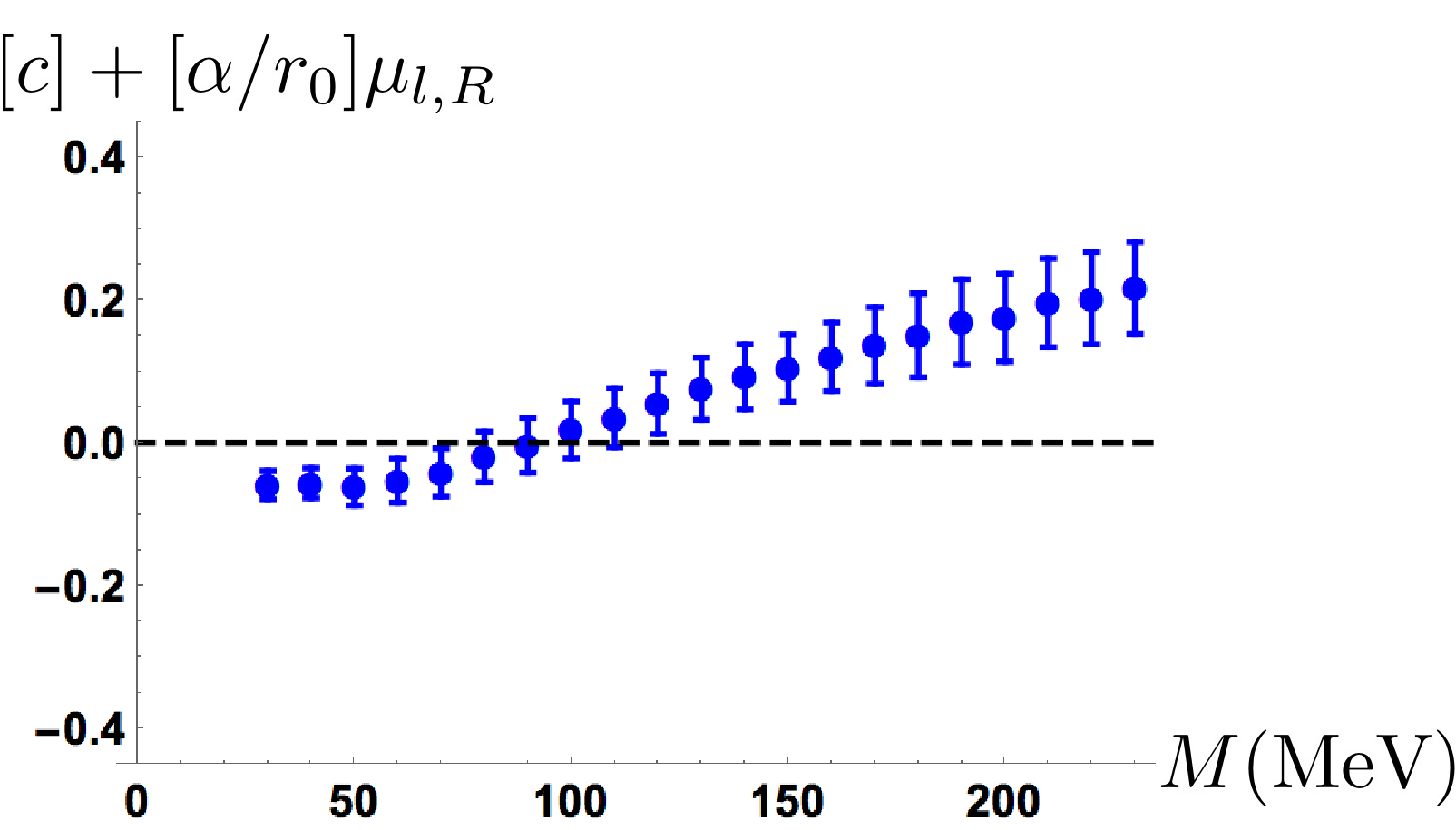}
   \caption{}
   \label{fig:ErrA}
\end{subfigure}
   \begin{subfigure}[t]{0.49\textwidth}
   \centering
   \includegraphics[width=\textwidth]{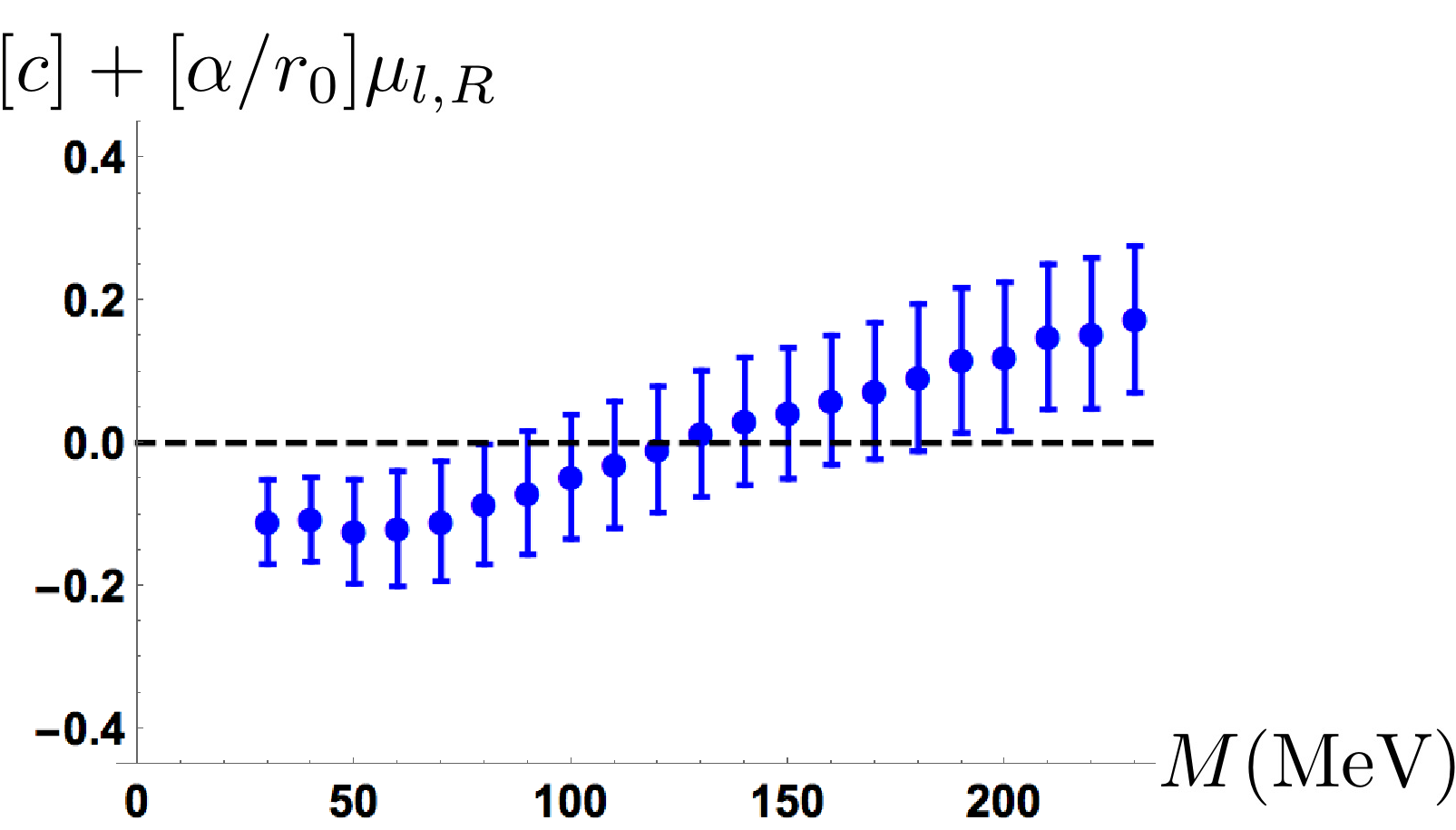}
   \caption{}
   \label{fig:ErrD}
   \end{subfigure}
   \begin{subfigure}[t]{0.49\textwidth}
   \centering
   \includegraphics[width=\textwidth]{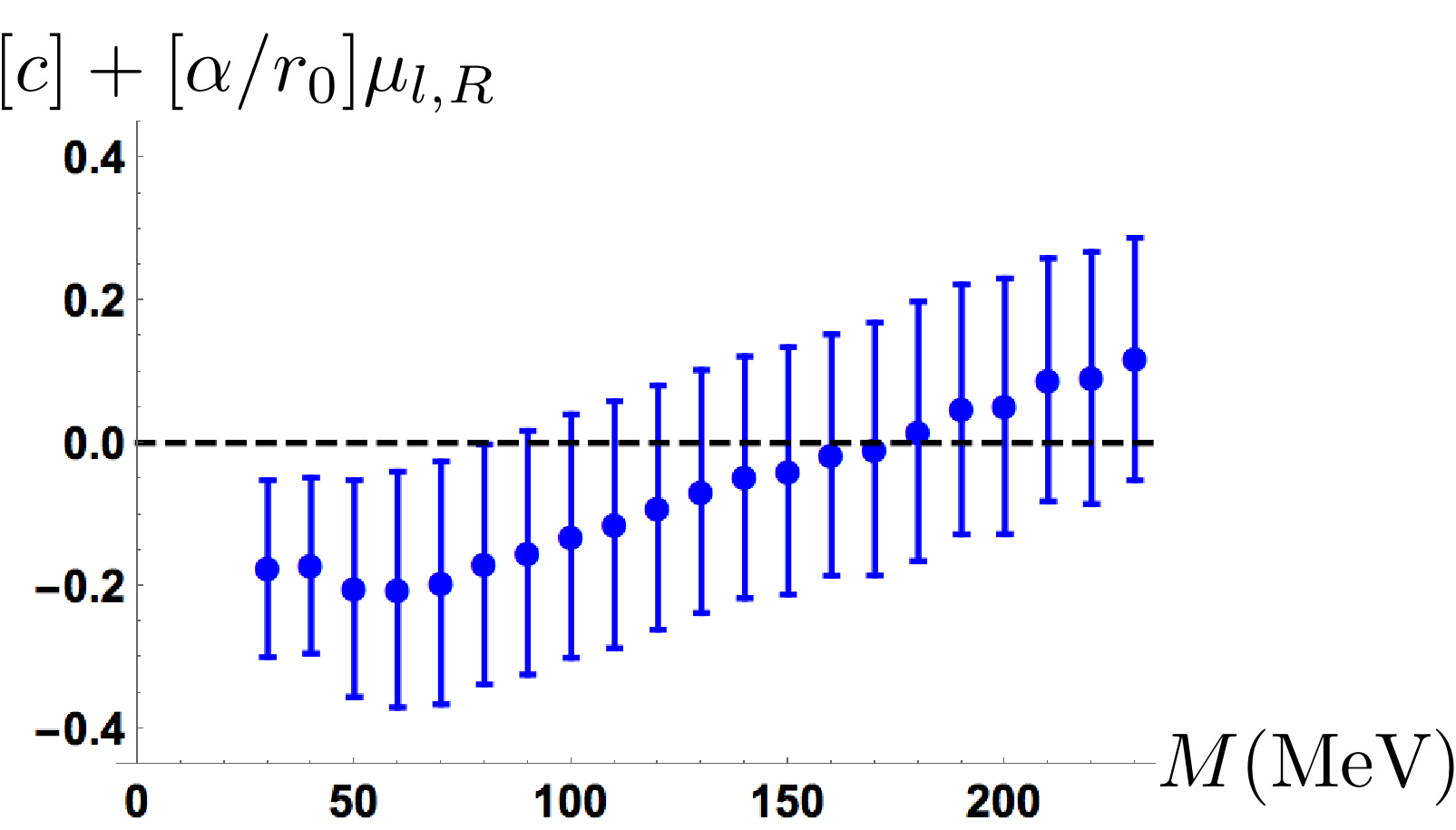}
   \caption{}
   \label{fig:ErrB}
   \end{subfigure}
      \caption{\small{Total discretization error from fitting to the form Eq.~(\ref{eqn:paramfit}) for $\mu_{l,R} = 12.6, 18.5, 26.0$ MeV, respectively. The data were calculated using the spectral projector definition of susceptibility seen in Eq.~(\ref{eqn:specproj}) with an $M$ corresponding to the value on the horizontal axis. The horizontal position of the black point is arbitrary and unrelated to the scale. Error bars are the result of 1000 bootstrap samples for each ensemble. Ensembles used are summarized in Table~\ref{tab:latticeparams1} and~\ref{tab:latticeparams2}. Fit parameters are shown in Table~\ref{tab:paramfit}.}}
\label{fig:TotalErr}
\end{figure}

One approach for determining the chiral condensate is a global 3-parameter fit to data at different 
values of the quark masses and the lattice spacing. Specifically, we fit to the expression:
\begin{equation}
r_0^4 \chi = \left[r_0^3 \Sigma\right] \,\frac{r_0 \mu_{l,R}}{2} + \left[c\right] \left(\frac{a}{r_0}\right)^2 + \left[\frac{\alpha}{r_0}\right] \,r_0 \mu_{l,R}\, \left(\frac{a}{r_0}\right)^2,
\label{eqn:paramfit}
\end{equation}
where $r_0$ is the Sommer parameter, used to make all quantities dimensionless, $\Sigma$ is the renormalized chiral condensate, and $c$ and $\alpha$ are the coefficients of the $\mathcal{O}(a^2)$ and $\mathcal{O}(\mu_l a^2)$ discretization terms. 
The square brackets indicate the actual fitting parameters.
Other expressions were tested, including a two parameter fit, where $\alpha$ was set to zero. 
Ultimately, the three parameter fit in Eq.~(\ref{eqn:paramfit}) was chosen, as it consistently had the smallest $\chi^2$ per degree of freedom for each value of $M$.
Cross sections of the fit at different lattice spacings are shown in Fig.~\ref{fig:paramA},~\ref{fig:paramB}, and~\ref{fig:paramD} for three values of $M$ and the GF definition. The fits themselves, for each of these values of $M$ and the GF, are shown in Fig.~\ref{fig:param80},~\ref{fig:param120},~\ref{fig:param160}, and~\ref{fig:paramGF}. 
In Fig.~\ref{fig:paramSig} extracted values of $r_0 \Sigma^{1/3}$ are shown for a large range of values of $M$ and for the gluonic definition. The total discretization coefficient, $[c] + [\alpha / r_0] r_0 \mu_{l,R}$, is shown for three values of $\mu_{l,R}$ in Fig.~\ref{fig:TotalErr}.
A summary of the obtained fit parameters is presented in Table~\ref{tab:paramfit}.

The first observation from Fig.~\ref{fig:paramSig} is that a consistent value of $\Sigma^{1/3}$ is obtained for any value of $M$ in the spectral projector method, with a small tendency for a smaller value for $M$ below around $50$ MeV.
The value from the gluonic GF definition is consistent with all the spectral projector values. 
The statistical error from the spectral projector definition is a factor of 2-4 smaller than the one from the gluonic definition. 
It should be noted that the statistical error noticeably increases for $M\gtrsim170$ MeV due to fewer ensembles being included in the fits.

We also observe that the spectral projector definition produces a topological susceptibility with much smaller discretization effects compared to the gluonic definition, regardless of the choice of $M$. 
By looking at Fig.~\ref{fig:TotalErr}, it can be observed that value of $M$ which minimizes discretization effects has some quark mass dependence. 
For the smallest mass, $\mu_{l,R}=12.6$ MeV, the optimal value is at $\sim 90$ MeV (with cut-off effects being compatible with zero in the range $70-120$ MeV). 
For $\mu_{l,R}=18.5$ MeV, $M$ is optimal at around $120$ MeV (compatible with zero for $80-180$ MeV). 
Finally, for the largest mass, $\mu_{l,R}=26.0$ MeV, $M$ is optimal at $\sim 170$ MeV (compatible with zero for $80-230$ MeV). 

It appears that the spectral projector method has larger mass dependent discretization effects compared to the gluonic definition, as $\alpha/r_0$ has larger magnitude, and smaller mass independent discretization effects, as $c$ is smaller. 
It should be noted, however, that the statistical error in $\alpha/r_0$ is large enough that the values for both spectral projectors and the gluonic definition are compatible. 

~

\begin{table}
  \centering
  \begin{tabular}{cccccccc}
    \hline
  \hline
   $M$ [MeV] & $\Sigma^{1/3}$ [MeV] & $r_0 \Sigma^{1/3}$ & $c$ & $\alpha/r_0$ & $N_{ens}$ & $\chi^2/d.o.f.$\\
\hline
80 & 316(18) &0.760(44)&0.121(32)&-4.7(1.4)& 9 & 0.281 \\
120 & 318(21) &0.764(50)&0.191(46)&-4.6(1.8)& 9 & 0.233 \\
160 & 318(23)&0.764(55)&0.249(54)&-4.3(2.0)& 9 & 0.198 \\
200 & 318(30) &0.764(73)&0.291(60)&-3.9(2.5)& 6 & 0.066 \\
GF & 289(49) &0.69(12)&1.15(12)&0.7(3.8)& 9 & 0.261\\
  \hline
  \hline
\end{tabular}
  \caption{\small{Fit parameters from Eq.~(\ref{eqn:paramfit}) and quality of the fit, $\chi^2/d.o.f.$. $M=80-160$ MeV are global fits using all ensembles in Table \ref{tab:latticeparams1} (apart from the ones in bold font). $M=200$ MeV excludes ensembles A30.32, D20.48, and D30.48, as 400 eigenmodes were not enough to reach $M=200$ MeV for these ensembles. Statistical errors are the result of a bootstrap procedure with 1000 samples and appropriate blocking.
\label{tab:paramfit}
}}
\end{table}

\subsection{$\chi(a\rightarrow0)$ LO Fit}
\label{subsec:extrap}

In order to obtain an estimate of a systematic error resulting from the choice of the ansatz for the fit, we have used a second method for extracting the continuum values of the chiral condensate. 
While each ensemble used has a different quark mass, each lattice spacing has a small, medium, and large quark mass ensemble. 
First, we choose one representative small, medium, and large quark mass and find the comparable values of topological susceptibility for each lattice spacing by performing a small linear interpolation to the same mass. For example, for the small quark mass, we choose $\mu_{l,R}=12.6$ MeV, which is the mass of the A30.32 ensemble, seen in Table~\ref{tab:latticeparams1}. 
For the B ensembles, for which $a\approx0.082$ fm, the $\mu_{l,R}=12.6$ MeV topological susceptibility was found by linearly interpolating between the B25.32 and B35.32 values, $12.3$ MeV and $18.5$ MeV, respectively.

The continuum limit of the topological susceptibility is taken for these three masses and obtained by performing a fit linear in $(a/r_0)^2$, using the ansatz:
\begin{equation}
r_0^4 \chi (a) = [K] (a/r_0)^2 + [r_0^4 \chi (a=0)],
\label{eqn:M2}
\end{equation}
where $r_0^4 \chi (a=0)$ is the continuum value of the susceptibility. 
The square brackets again denote the actual fit parameters.
The extracted values of the susceptibility can be found in Table~\ref{tab:M2}. 
\begin{table}
   \centering
  \begin{tabular}{ccccc}
      \hline
          \hline
    $\mu_{l,R}$ [MeV] & $M$ [MeV] &  $r_0^4 \chi (a=0)$ & $K$ \\
    \hline
    12.6 & 80 & 0.0088(16) & -0.084(48)\\
    12.6 & 120& 0.0093(20)& -0.020(62)\\
    12.6 & 160& 0.0096(23)& 0.032(71)\\
    12.6 & GF& 0.0075(38)& 1.11(13)\\
    18.5 & 80& 0.0120(13) & -0.166(48)\\
    18.5 & 120& 0.0121(15)& -0.079(60)\\
    18.5 & 160& 0.0118(16)& -0.003(68)\\
    18.5 & GF& 0.0088(28)& 1.10(13)\\
     26.0 & 80& 0.0143(24)& -0.191(77)\\
    26.0 & 120& 0.0151(29)& -0.124(94)\\
    26.0 & 160& 0.0147(32)& -0.05(10)\\
    26.0 & GF& 0.0101(60)& 1.22(20)\\
    \hline
        \hline
\end{tabular}
  \caption{\small{Fit parameters from Eq.~(\ref{eqn:M2}) using ensembles from Table \ref{tab:latticeparams1}. Values labeled with $M$ were calculated using the spectral projector definition of susceptibility seen in Eq.~(\ref{eqn:specproj}). Values labeled with ``GF" were calculated using a gluonic definition seen in Eq.~(\ref{eqn:gfdef}) along with gradient flow. Statistical errors are the result of 1000 bootstrap samples for each ensemble.
}}
  \label{tab:M2}
\end{table}

\begin{figure}[t!]
   \centering
   \begin{subfigure}[t]{0.49\textwidth}
   \centering
   \includegraphics[width=\textwidth]{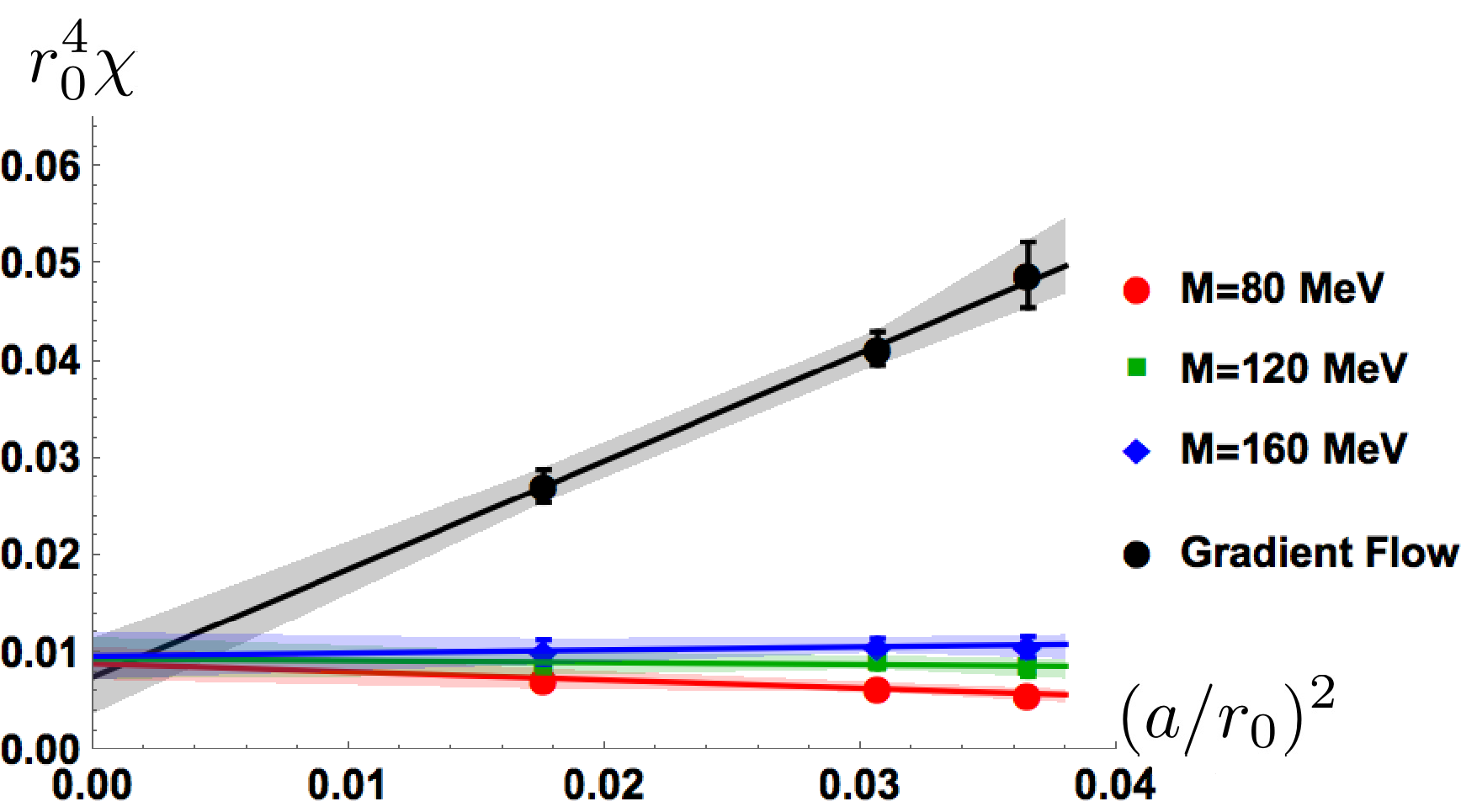}
   \caption{}
   \label{fig:Aextrap}
\end{subfigure}
\begin{subfigure}[t]{0.49\textwidth}
   \centering
   \includegraphics[width=\textwidth]{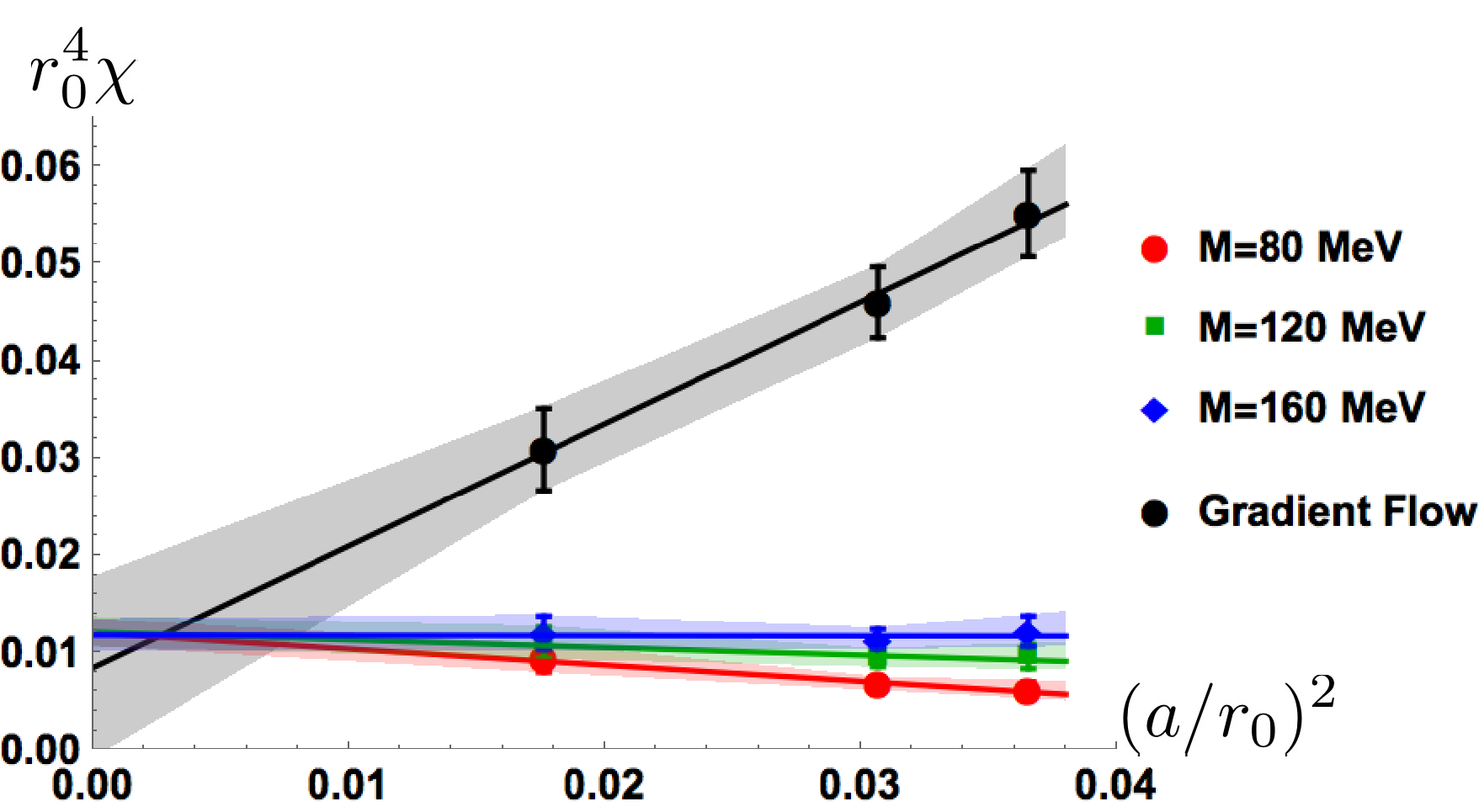}
   \caption{}
   \label{fig:Dextrap}
\end{subfigure}
\begin{subfigure}[t]{0.49\textwidth}
   \centering
   \includegraphics[width=\textwidth]{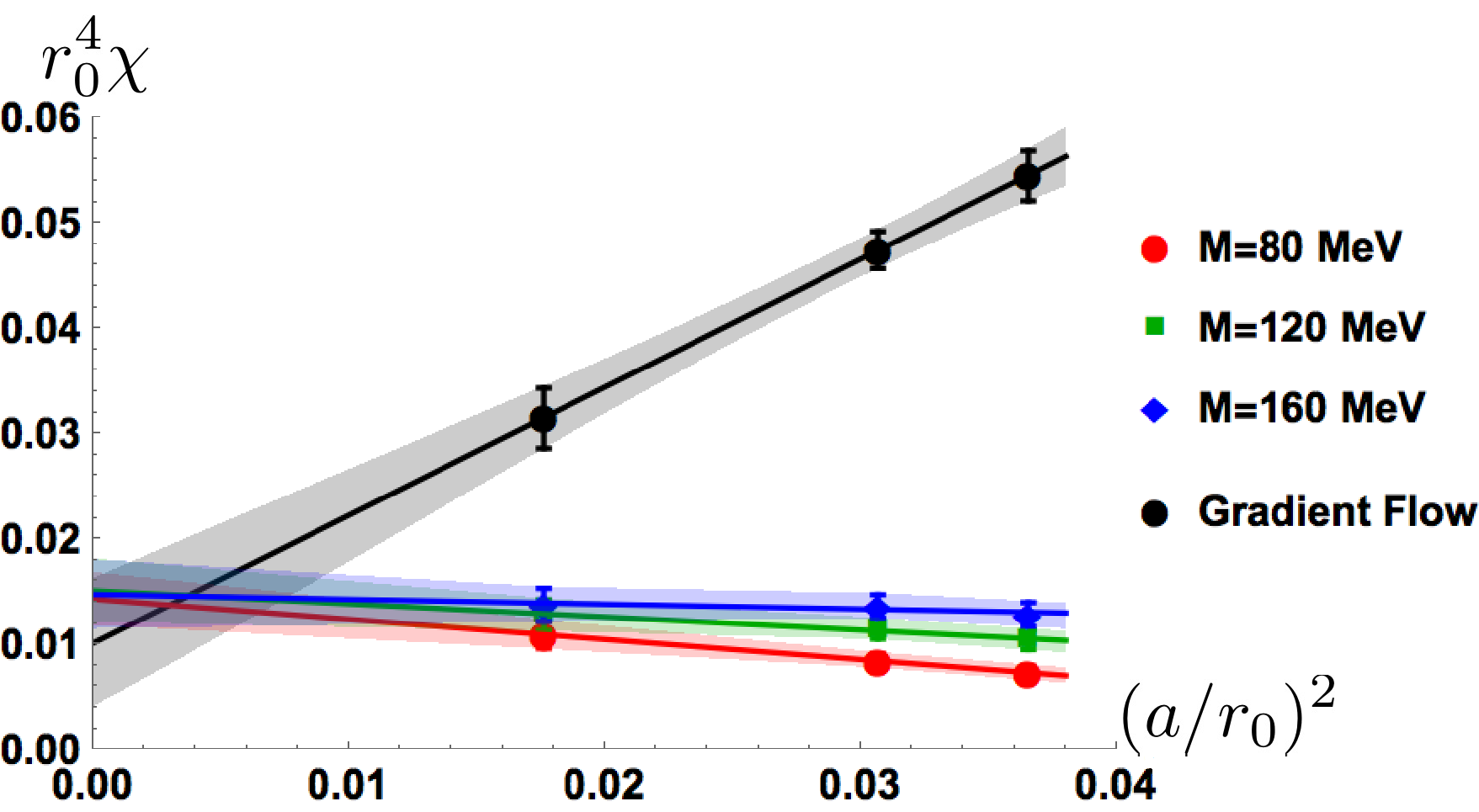}
   \caption{}
   \label{fig:Bextrap}
\end{subfigure}
   \caption{\small{Continuum extrapolations at $\mu_{l,R}=12.6, 18.5, 26.0$ MeV, respectively. The actual calculated value of the topological susceptibility was used for the largest lattice spacing, the other two were generated by a linear interpolation of results from the ensembles with the two nearest quark masses. 
   Statistical errors are the result of a bootstrap procedure with 1000 samples and appropriate blocking. 
   Ensembles used are summarized in Table~\ref{tab:latticeparams1}. 
   }}
\end{figure}

\begin{figure}[t!]
   \centering
   \begin{subfigure}[t]{0.49\textwidth}
   \centering
   \includegraphics[width=\textwidth]{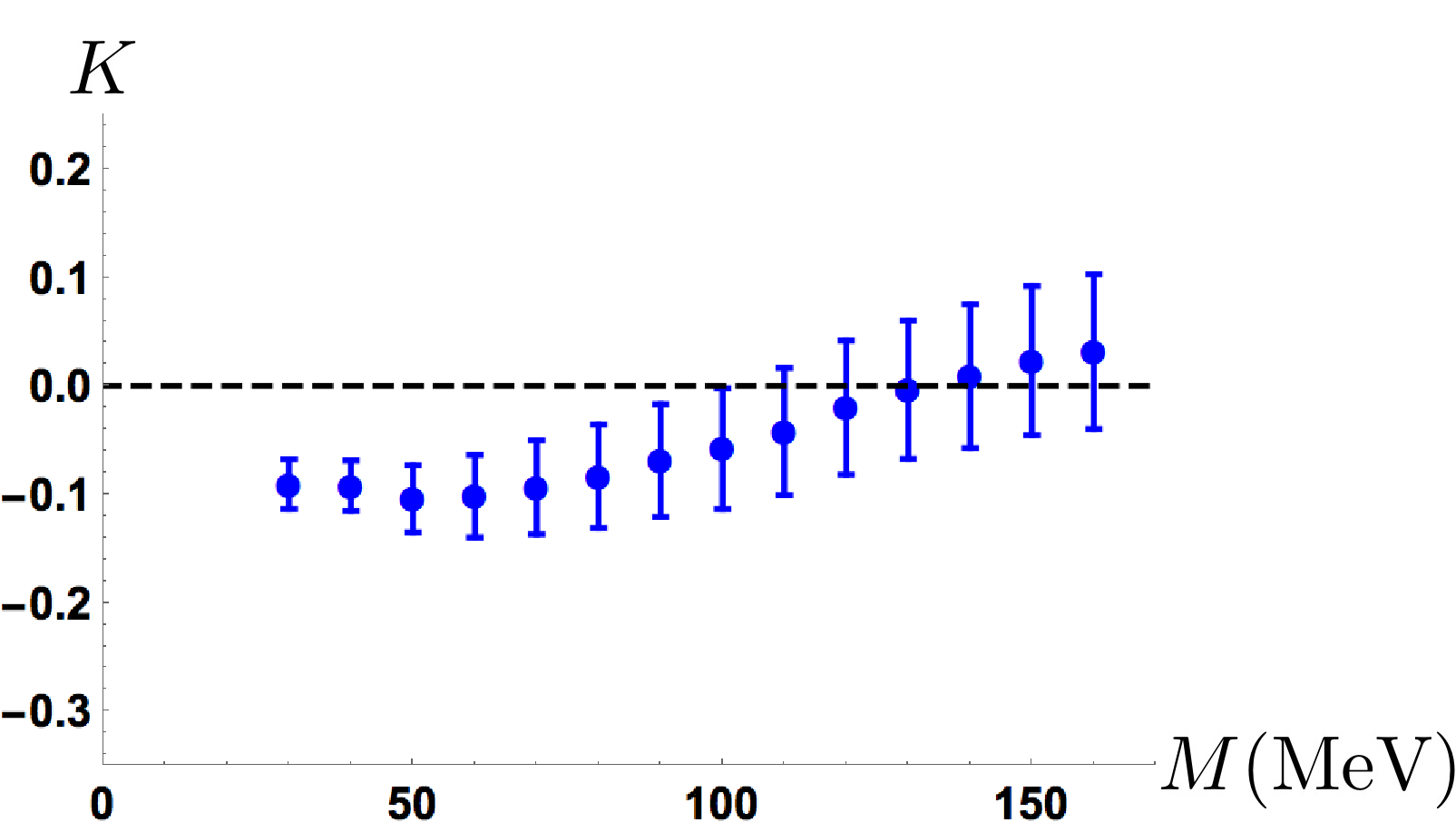}
   \caption{}
   \label{fig:ErrA}
\end{subfigure}
   \begin{subfigure}[t]{0.49\textwidth}
   \centering
   \includegraphics[width=\textwidth]{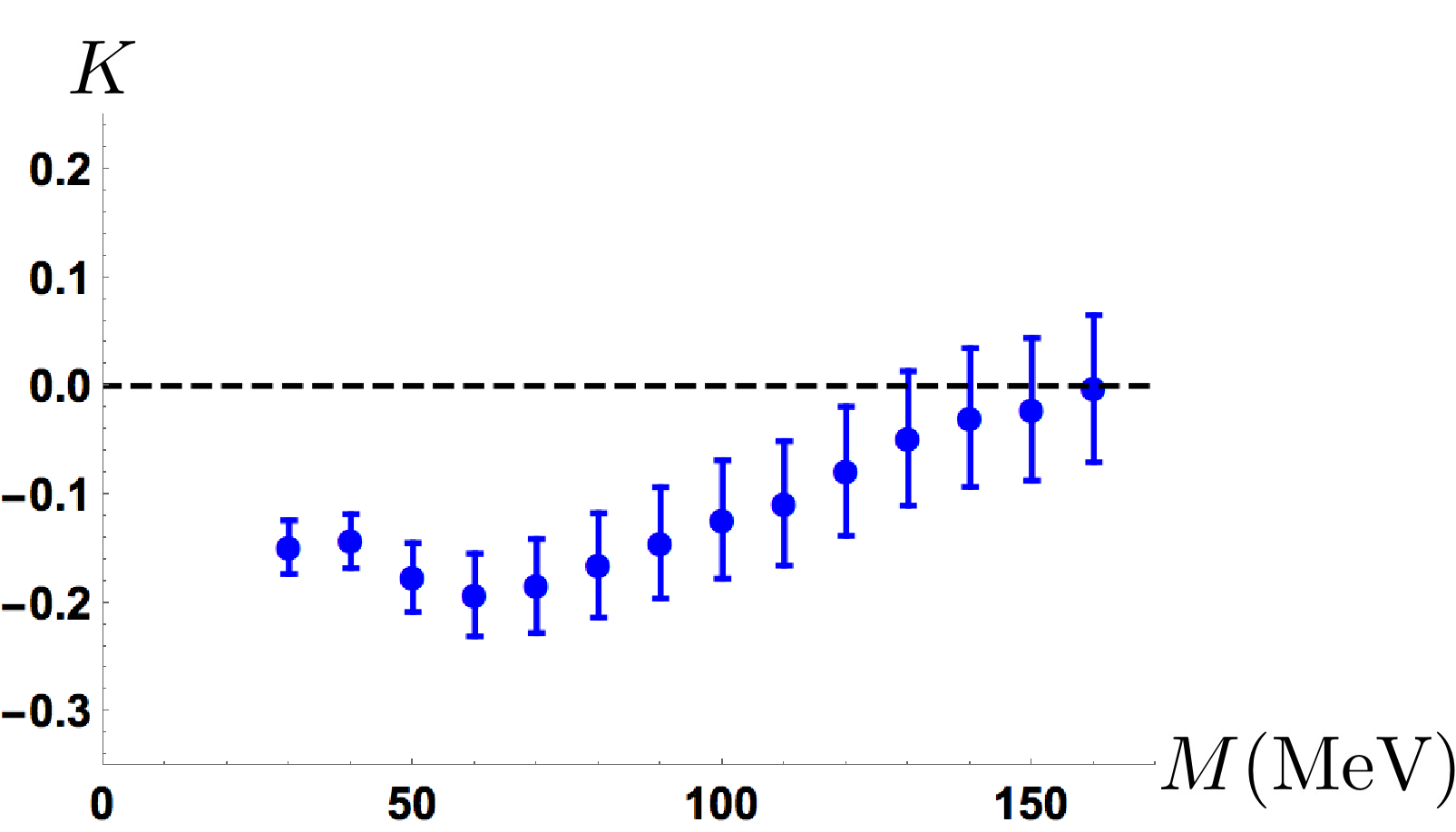}
   \caption{}
   \label{fig:ErrD}
   \end{subfigure}
   \begin{subfigure}[t]{0.49\textwidth}
   \centering
   \includegraphics[width=\textwidth]{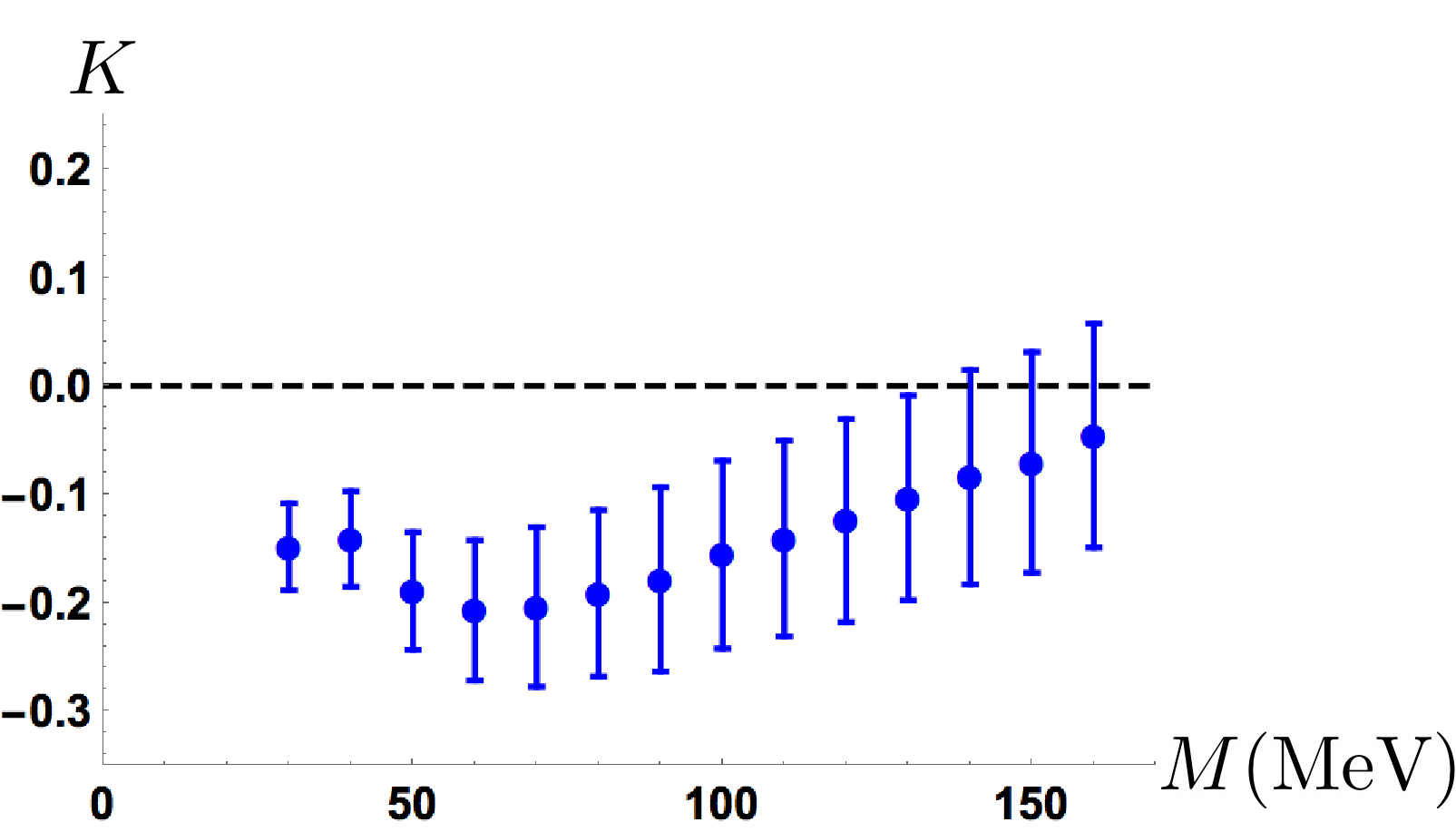}
   \caption{}
   \label{fig:ErrB}
   \end{subfigure}
      \caption{\small{Total discretization error from a fit to the form of Eq.~(\ref{eqn:M2}) for $\mu_{l,R} = 12.6, 18.5, 26.0$ MeV, in the plots (a)-(c), respectively. 
      The data were calculated using the spectral projector definition with an $M$ corresponding to the value on the horizontal axis.
      Statistical errors are the result of a bootstrap procedure with 1000 samples and appropriate blocking. 
      Ensembles used are summarized in Table~\ref{tab:latticeparams1}. 
      The resulting fit parameter, $K$, is shown in Table~\ref{tab:M2}.}}
\label{fig:M2Err}
\end{figure}

The results of this procedure are collected in Figs.~\ref{fig:Aextrap}, \ref{fig:Dextrap}, and \ref{fig:Bextrap}, 
for the small, medium, and large quark masses, respectively. 
A comparison of the discretization effects, contained in the parameter $K$, is shown for a range of values of $M$ in Fig.~\ref{fig:M2Err}.

\begin{figure}[ht!]
   \centering
   \includegraphics[width=.49\textwidth]{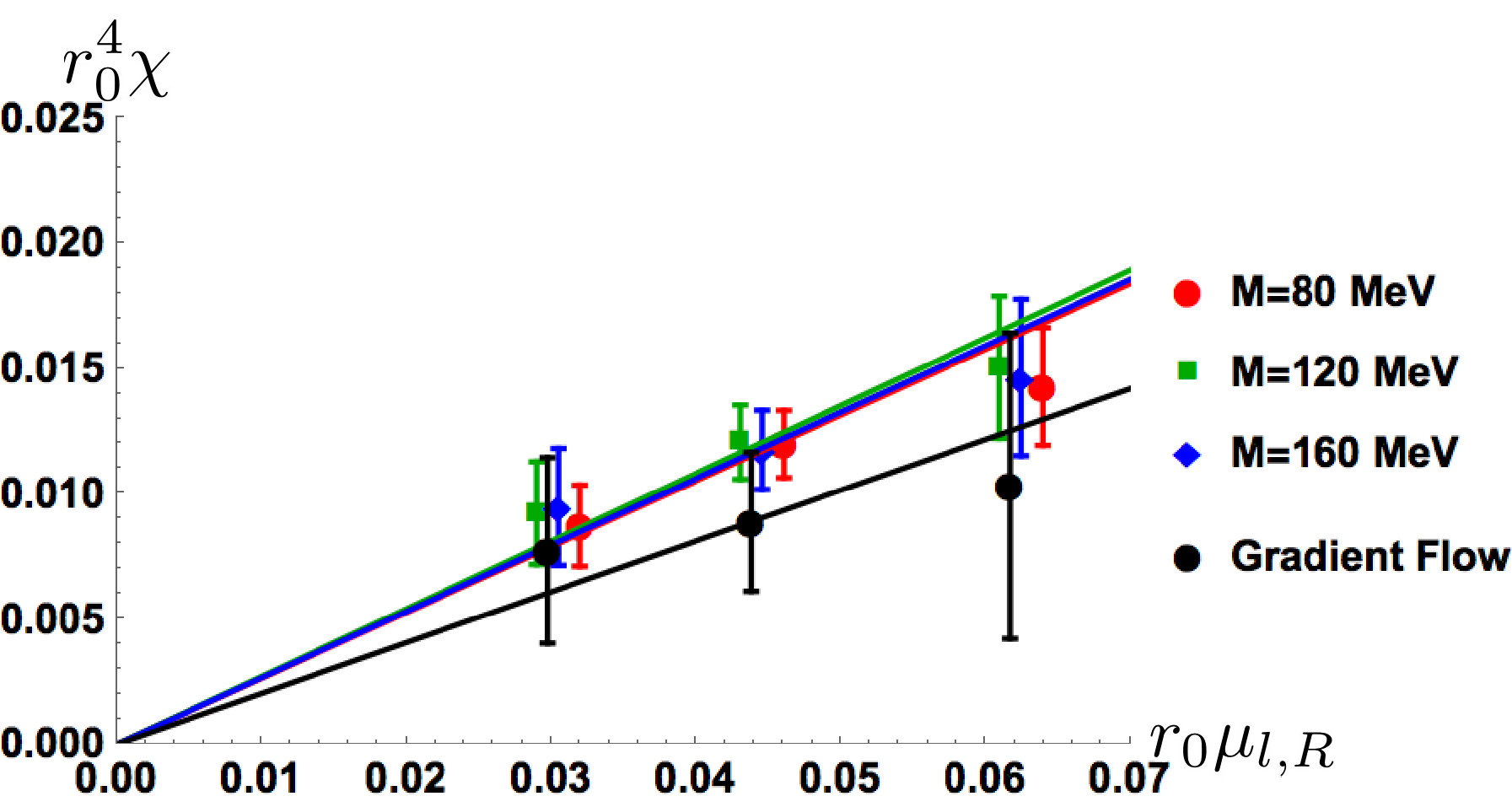}
   \caption{\small{Continuum fit to the lowest order $\chpt$ expression for topological susceptibility, Eq.~(\ref{eqn:extrapfit}). 
   Values are the continuum extrapolated ones seen in Figs.~\ref{fig:Aextrap},~\ref{fig:Dextrap}, and~\ref{fig:Bextrap}.
   Statistical errors are the result of a bootstrap procedure with 1000 samples and appropriate blocking. 
   Note that the values are offset on the horizontal axis for visibility.}}
   \label{fig:extrapfit}
\end{figure}

To calculate the chiral condensate, each extracted continuum value of the topological susceptibility is plotted against the quark mass and fitted to the continuum lowest order $\chpt$ formula Eq.~(\ref{eqn:contchi}), or in dimensionless form:
\begin{equation}
r_0^4\chi(a=0)=[r_0^3\Sigma] \frac{r_0 \mu_{l,R}}{2}.
\label{eqn:extrapfit}
\end{equation}
This fit is shown in Fig.~\ref{fig:extrapfit} and the corresponding values of the renormalized chiral condensate are presented in 
Table~\ref{tab:extrapfit}.

From Figs.~\ref{fig:Aextrap}, \ref{fig:Dextrap}, and \ref{fig:Bextrap}, it is apparent that for a given quark mass all values of $M$ and the gluonic definition converge to the same value in the continuum limit, within errors. This is as expected, as the definitions should only differ at finite lattice spacing.

We observe that the spectral projector definition produces a topological susceptibility with much smaller discretization effects compared to the gluonic definition, regardless of the choice of $M$. 
From Fig.~\ref{fig:M2Err} the optimal choice of $M$ in order to minimize discretization effects can be concluded. 
For the smallest mass, $\mu_{l,R}=12.6$ MeV, the optimal value is $\sim 130$ MeV (compatible with zero for $100-160+$ MeV). 
For $\mu_{l,R}=18.5$ MeV,  $M$ is optimal at $160$ MeV (compatible with zero for $120-160+$ MeV). 
For the largest mass, $\mu_{l,R}=26.0$ MeV, $M$ appears to be optimal at a value larger than, but near $160$ MeV (compatible with zero for $120-160+$ MeV). 
These ranges are consistent with the ones found from the global fit method, but are consistently slightly larger and have less mass dependence. 
Regardless of the method used, the spectral projector method for any $M$ has significantly smaller discretization effects compared to the gluonic definition.

The fit values of $r_0 \Sigma^{1/3}$ seen in Table~\ref{tab:extrapfit} are larger but within errors of the values found using the global fit (cf.\ Table~\ref{tab:paramfit}).
The statistical errors are smaller, likely due to being more constrained, only fitting to one parameter. However, the $\chi^2$ per degree of freedom is larger for the spectral projectors results, for which reason we prefer the values from the global fit.

\begin{table}[t!]
  \centering
  \begin{tabular}{cccccc}
  \hline
  \hline
   $M$ [MeV] & $\Sigma^{1/3}$ [MeV] & $r_0 \Sigma^{1/3}$ & $\chi^2/d.o.f.$\\
\hline
80 & 336(10) &0.807(23)& 0.555  \\
120 & 339(11) &0.815(26)& 0.344 \\
160 & 337(12) &0.809(29)& 0.387\\
GF & 308(26) &0.740(62)& 0.171\\
\hline
\hline
\end{tabular}
  \caption{\small{Values of the renormalized chiral condensate, from fits of Eq.~(\ref{eqn:extrapfit}) and quality of the fit, $\chi^2/d.o.f.$. These values are the result from a fit to extracted continuum values of the renormalized chiral condensate, using the method described in Sec.~\ref{subsec:extrap}. 
  Statistical errors are the result of a bootstrap procedure with 1000 samples and appropriate blocking. 
      }}
\label{tab:extrapfit}
\end{table}

\section{Conclusions}
\label{sec:conclusions}

In this paper, we have compared the calculation of the topological susceptibility and the chiral condensate using the spectral projector method to an $\mathcal{O}(a^2)$-improved gluonic definition using the gradient flow.
The topological susceptibility computed using spectral projectors was found to have much smaller discretization effects in the considered setup. 
It should be noted, however, that in the case of the gradient flow an improved definition could be derived which reduces cut-off effects significantly~\cite{Ramos:2014kka}. It would therefore be interesting, whether such an improved definition can also be found for the here considered topological susceptibility. 
We have also determined that a spectral cutoff as small as $M\sim30$ MeV is sufficient for extracting fit quantities, such as the chiral condensate. 
The optimal choice of $M$ which minimizes discretization effects depends on the details of the ensemble, specifically the quark mass, with lighter masses corresponding to smaller $M$. 
In this investigation, we see from Fig.~\ref{fig:paramSig} that our choice of $M$ has little effect on $r_0\Sigma^{1/3}$, as long as discretization effects are understood and parametrized. For citing a value, we use $M=120$ MeV, as it is in the range of the optimal value of $M$ for most of our estimated ranges. 
Taking the result from the global fit from Table~\ref{tab:paramfit} as the central value and the difference with respect to the result from the continuum extrapolated susceptibility as the estimate of systematic error, we finally quote:
\begin{equation}
r_0 \Sigma^{1/3} = 0.764(50)_{\rm stat}(51)_{\rm sys}, \quad \Sigma^{1/3} = 318(21)_{\rm stat}(21)_{\rm sys}~{\rm MeV}. 
\end{equation}
Where we have used $r_0=0.474~{\rm fm}$~\cite{Carrasco:2014cwa} to convert to physical units. Comparing to our previously calculated values, this result is slightly larger. One value, $r_0 \Sigma^{1/3}=0.651(61)$ \cite{Cichy:2013rra} is from a fit of the leading order $\chpt$ to the quark mass dependence of the topological susceptibility evaluated with stochastic spectral projectors on a similar set of ensembles. Another, $r_0 \Sigma^{1/3}=0.689(16)(29)$ \cite{Cichy:2013gja} resulted from direct extraction of the condensate from the mode number evaluated also with stochastic spectral projectors on a similar set of ensembles. While slightly larger, this result is still compatible within errors.

By choosing an appropriate value of $M$, we have shown the method of spectral projectors eliminates discretization artifacts in topological susceptibility up to our computed percision. This is highly relevant for calculating topological charge dependent quantities, when there is only one lattice spacing available and a continuum extrapolation cannot be performed. 
This is at present the case with the physical point ensembles recently generated by the ETMC~\cite{Alexandrou:2017qyt,Abdel-Rehim:2015pwa,Abdel-Rehim:2015owa}. 
We believe that this approach will significantly reduce discretization errors due to the topological charge, while preserving reasonable statistical errors, allowing more accurate computation of experimentally relevant physical parameters.

\section{Acknowledgements}

We thank the European Twisted Mass Collaboration for generating ensembles of gauge field configurations that we have used for this work and for a very enjoyable collaboration. We acknowledge helpful discussions with Dean Howarth on technical details.
K.C.\ was supported in part by the Deutsche Forschungsgemeinschaft (DFG), project nr. CI 236/1-1. This work used computational resources from the Swiss National Supercomputing Centre (CSCS) under project IDs s540, s625 and s702, from the John von Neumann-Institute for Computing on the Jureca and the BlueGene/Q Juqueen systems at the research center in Juelich, with project ID ECY00, and from Gauss on SuperMUC with allocation ID 44060. C.L. would like to acknowledge supported from the PRACE-4IP Summer of HPC program under EC grant agreement number 653838. A.A. has been supported by an internal program of the University of Cyprus under the name of BARYONS.

\bibliographystyle{h-physrev}
\bibliography{CondensatePaper}

\end{document}